# A primer on correlation-based dimension reduction methods for multi-omics analysis

*Running head: Correlation-based methods for multi-omics*

**Author information**
Tim Downing[1,2] (Orcid: 0000-0002-8385-6730), Nicos Angelopoulos[1] (Orcid: 0000-0002-7507-9177).
[1] Pirbright Institute, Surrey, UK.
[2] Dublin City University, Dublin, Ireland.
Corresponding author: Tim.Downing@pirbright.ac.uk.

**Abstract**

The continuing advances of omic technologies mean that it is now more tangible to measure the numerous features collectively reflecting the molecular properties of a sample. When multiple omic methods are used, statistical and computational approaches can exploit these large, connected profiles. Multi-omics is the integration of different omic data sources from the same biological sample. In this review, we focus on correlation-based dimension reduction approaches for single omic datasets, followed by methods for pairs of omics datasets, before detailing further techniques for three or more omic datasets. We also briefly detail network methods when three or more omic datasets are available and which complement correlation-oriented tools. To aid readers new to this area, these are all linked to relevant R packages that can implement these procedures. Finally, we discuss scenarios of experimental design and present road maps that simplify the selection of appropriate analysis methods. This review will guide researchers navigate the emerging methods for multi-omics and help them integrate diverse omic datasets appropriately and embrace the opportunity of population multi-omics.

**Key messages:**
1. The early integration of multi-omic datasets is essential.
2. The dataset and research question complexity determine the analysis methods applied.
3. Optimal approaches should use a combination of supervised and unsupervised tools.

## Basic concepts in multi-omics

Combining important omics data features appropriately across the genome, regulome, methylome, transcriptome, translatome, proteome and interactome is challenging. These omic data types are complex, heterogeneous and possess high dimensionality. Genome, regulome, methylome, transcriptome and translatome data are typically generated using high-throughput sequencing of DNA/RNA that can be deciphered into a set of features across the samples sequenced (Table 1). Proteomic data is created using mass spectrometry of lysed samples separated by liquid chromatography, and interactome data is made from protein binding experiments such as chromatin immunoprecipitation sequencing (ChIP-Seq). For instance, viral infections have complex molecular patterns in hosts, ranging from acute disease to latent oncogenic process. Such an integrated approach could more effectively help us understand infection mechanisms and host responses by linking viral genome data to taxonomically classify the infecting lineage, RNA-Seq to assess the host immune changes over time across cell types, and ChIP-Seq to gather host-virus protein binding patterns.

| Technology | Biological data | Simplest data states | Example |
| --- | --- | --- | --- |
| Genomics | Mutations | Two or three discrete states | Sars-CoV-2 infection |
| Regulomics | Regulatory element location | Quasi-continuous | Identifying enhancer or promoter elements |
| Methylomics | Methyl marks | Quasi-continuous | Testing for gene regulation |
| Transcriptomics | Gene activity | Continuous | Host response to viral infection |
| Translatomics | mRNA translation rates | Quasi-continuous | Isoform translation and translation efficiency |
| Proteomics | Protein expression | Continuous | Host protein levels |
| Interactomics | Protein-DNA and protein-protein binding | Two discrete states | Host-virus infection mechanisms |

**Table 1**. The major omic technologies, their biological data type, data structures and simple examples of applications.

We can define multi-omics analysis as the early combination of at least two omic technologies where the data is integrated across features (Figure 1). The early integration of data is important: relying solely on post hoc conceptual integration after analysing separate omic datasets may inadvertently overlook crucial characteristics intrinsic to each individual technology [1]. This differs from conceptual, sample-clustering and concatenation-based approaches, which limit inferential power due to the *post hoc* nature of clustering many samples across linked omic datasets [2]. Concatenating disparate omic datasets together is biased due to the varied heterogeneity of the data types, their relative numbers of features, and their differing sources of error per omic type [3]. The main purpose of this review is to outline better approaches that allow early omic data integration. These may use matrix factorisation of the data together including aligned features (molecules), typically after transformation and/or scaling relevant to the technology (Figure 1). This can be followed by dimension reduction to get fewer features in the form of higher-level components. And next we can cluster and model across sample-component pairs, as well as the samples and components individually, to illuminate how the samples, features, technologies and their interactions related to one another. This modelling can also be applied by network-based methods following or replacing matrix factorisation.

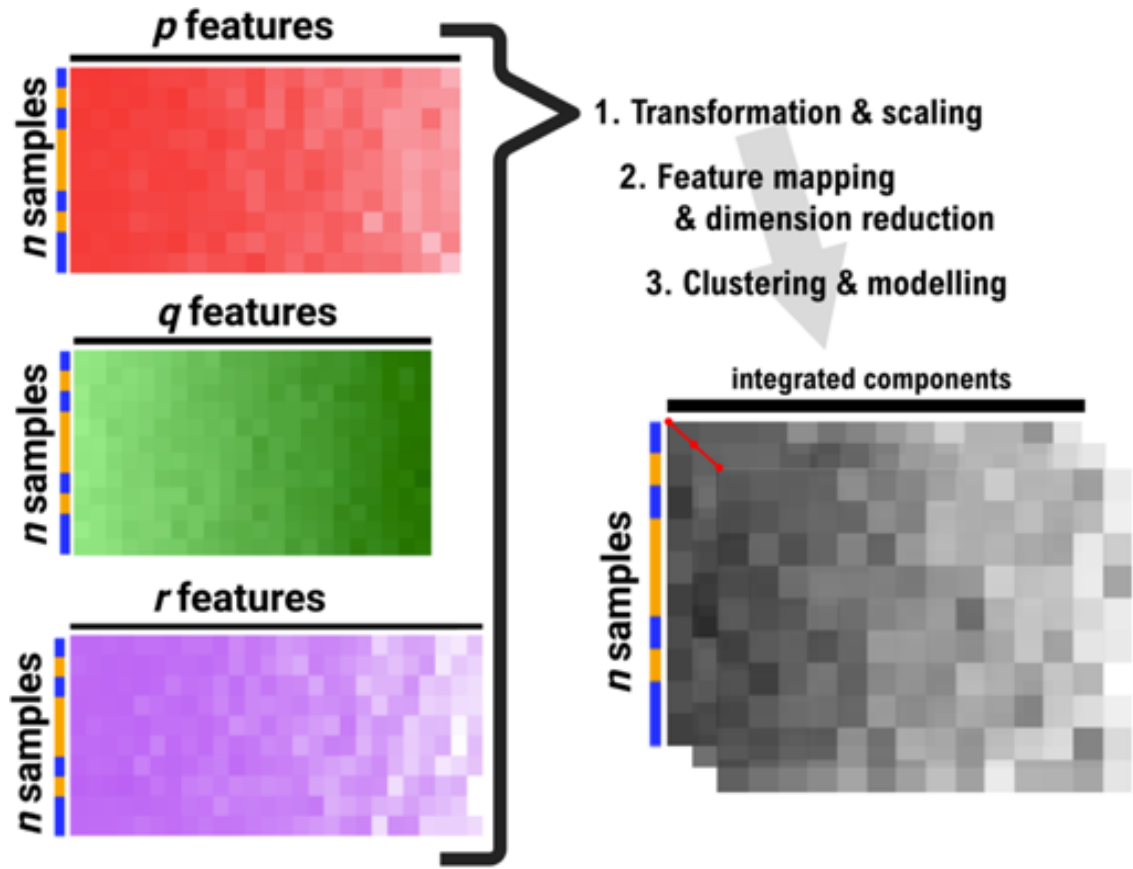

**Fig 1**. Schematic representing central concepts in multi-omics relevant to matrix factorisation. For three omic datasets generated for *n* samples containing *p* (red), *q* (green) and *r* (mauve) features respectively, multi-omic analysis involves (1) transforming this data to scale it within and across omic platforms; (2) mapping features between omic types and applying dimension reduction techniques to retain informative features (*k* components such that $r > p > q \gg k$) only; (3) clustering relevant samples and components and modelling the interactions between components or features. This integration creates a multi-dimensional dataset, represented by variation across samples that may have differing associated metadata (blue vs orange), *k* components (black) and omic type (red). The diagram shows three dimensions and extending this dataset to 4+ could be achieved by adding (say) a temporal dimension.

When multi-omics is applied to large numbers of related samples, it can decipher their molecular features and phenotypes [4]. This population multi-omics is the early combination of at least two omic technologies across features for a large biologically related sample collection. They could come from different cohorts of patients, or are genetically diverse isolates, or possess varied phenotypes, or *in vitro* work. Together, this systems approach seeks to understand the emergent properties in a set of samples and features that may not be evident without the early integration of extensive molecular data. Understanding virus-host interactions using data-driven approaches to mitigate heterogeneity and identify opaque associations has been achieved well for new emerging viruses like Sars-CoV-2 [5]: such approaches can be applied to other viral pathogens. In this review, we focus on concepts and working tools relevant for omics analyses through R: particularly for researchers with stronger backgrounds in biology relative to statistics. Extensive reviews of tools and models across other languages and oriented to human samples have been published elsewhere (e.g. [6]), and are beyond the scope of this paper.

**1 Correlation-based dimension reduction models for single omic datasets**

## 1.1 An overview of dimension reduction goals and approaches

Dimension reduction peels off layers of unnecessary complexity of multivariate data with many features by converting it into less complex data with fewer summarised features, while ideally retaining key properties of the original data. Such features could be gene expression values. We assume here that the summary patterns extracted (called components) provide insightful information. We also assume that each component correlates with a number of the original features. So, for an omic dataset with *n* samples and *p* features, we aim to obtain *k* components where k is much smaller than p (k << p). Depending on the approach, the original data may be transformed extensively and normalised during this process. In addition, most omic datasets are sparse, which means many features are less informative, so many dimension reduction methods possess sparse versions to facilitate feature selection where many features may have no information.

Many dimension reduction methods are based on the principle that the data can be transformed in linear space. Although many nonlinear methods exist, they have a wider spread of outcomes that depend on the dataset structure so they are less applicable here, given omic data's quantitative nature, can be potentially hard to interpret, and have been detailed elsewhere [7]. Hence, we focus on linear approaches here. For a single omic matrix *X* with *n* samples and data for *p* features ($X \in R^{n \times p}$), we can define each entry in that matrix across a total of *K* components with a Gaussian error term $\varepsilon_{nk}$ for sample *n* in component *k* and feature *p* as: $x_{np} = \sum_{k=1}^{K} w_{nk} h_{kp} + \varepsilon_{nk}$. Here, $w_{nk}$ is the score reflecting the sample's patterns, and $h_{kp}$ is the estimated transformed data, reflecting the feature's variation. In this way, matrix *X* can be estimated by a pair of matrices, *W* and *H*, reflecting the loadings and components, respectively.

Many dimension reduction methods have both supervised and unsupervised implementations. Supervised approaches require data groups or labels to be specified in advance, whereas unsupervised ones do not require any specific metadata or structure. Both can be used in tandem: unsupervised approaches for data exploration and quality control, and supervised methods for hypothesis-driven tests. Unsupervised approaches typically aim to maximise the fraction of total feature variation explained by the components. Supervised methods differ because they usually select components maximising differences between groups. Many supervised methods develop models based on training data that is tested on new data to assess model performance, and this can then feed back into parameter optimisation at the training stage. Consequently, the features and components detected by supervised and unsupervised analyses will differ. This means assessing them in tandem can be effective: the supervised methods examine the differences between known groups, unsupervised ones explore the level of variability present, and true effects will show congruence in their combined results.

## 1.2 Feature extraction

Dimension reduction methods are usually feature extraction methods: the most useful in omics analyses being linear transformations because the retention of linear biological signals allows better and more accurate interpretation. Feature extraction is sometimes termed feature projection. To make this review accessible for readers, we generalise by discussing methods that use either correlation- or projection-based feature extraction. Both aim to reduce the dimensionality of the input datasets. Projection-based methods aim to capture the important relationships between features using a new subspace, whereas correlation-based approaches focus on the features' covariances to eliminate uninformative ones. The delineation between correlation- or projection-based is sometimes unclear because the latter often use correlation metrics to aggregate features into components that define the new subspaces.

Most methods outlined here have a single global minimum (for a review of non-convex methods, see [7]). These typically use Euclidean distances in matrix eigendecomposition to obtain the covariances across features or similarities across samples because Euclidean distances are robust measures [8]. Feature extraction summarises a set of features as a component (or latent factor). It can also identify the original features that gave rise to a component. There are several methods of feature extraction and feature selection. Factor analysis is a form of feature extraction that reduces the numbers of observed features to a smaller number of latent variables termed factors such that the data can be explained by a combination of these factors, with some error.

Feature selection differs from feature extraction because it directly samples from the initial features, rather than abstracting them. Filtration of uninformative features is an example of feature selection. Filter-based (also called relief-based) approaches are the most common when dealing with omic data: these examine associations between features based on correlations, distances or information content. Other approaches include embedded and wrapper-based methods. Wrapper-based methods produce a model that starts with all features included, and then the model performance is evaluated as features are gradually removed, so that selected features are retained only if they improve or have marginal negative effect on the predictive power of the model.

Matrix factorisation is a type of unsupervised projection-based method which decomposes the input matrix based on combinations of linear models. It includes principal component analysis (PCA), non-negative matrix factorisation (NMF) and other approaches. Many matrix factorisation methods use singular value decomposition (SVD) or an equivalent. SVD is a generalisation of eigendecomposition to a rectangular matrix. If we have a dataset of *n* samples and *p* features (denoted as $X \in R^{n \times p}$). This produces eigenvalues, which reflect the magnitudes of the *k* PCs in descending order, and it creates eigenvectors, which denote the samples' coordinates in each PC.

Other dimension reduction methods attempt to decompose (or factorise) matrix *X* into two matrices across *k* components, one reflecting the samples ($W \in R^{n \times k}$) and the other the features ($H \in R^{k \times p}$) as *X~WH*. $W_{n \times k}$ is called the loadings and it also referred to in the literature as the source matrix or features matrix. $H_{k \times p}$ is the components, which can also be termed the coefficients matrix latent variables, latent factors, score matrix, rotation matrix, unmixing matrix, weights matrix, signatures, eigengenes, or meta-genes. Many dimension reduction methods iteratively minimise the error between *X* and *WH* using a variety of algorithms/approaches.

**1.3 Principal component analysis (PCA)**

PCA is an unsupervised linear dimension reduction method that identifies correlations across features based on their effect on samples. It maximises the effects of features distinguishing between samples. The eigenvectors (principal components, PCs) represent data from a range of input features [9]. It can be carried out via SVD or eigendecomposition of the data covariance matrix, and calculates the samples' coordinates (eigenvectors) across orthogonal PCs determined from the features, where the PCs are ordered by the fraction of total variation explained. This is applied iteratively such that each additional PC is independent (orthogonal) of the previous ones and the error is minimised. The scaling factors chosen for the original features are designed to maximise the variation explained by each PC. R functions for PCA (and MDS) are part of the R base functions [10] (Table 2). The input data for PCA should not have extreme outliers, be unimodal, and have linear effects. If these assumptions are not met, PCA plots may show horseshoe or arch effects, and alternative approaches should be applied [11].

| Name | Full method name | Package | function(s) | Reference |
|---|---|---|---|---|
| PCA | Principal Component Analysis (PCA) | base R | prcomp, princomp | [12] |
| | | dimRed | PCA | [129] |
| | | mixOmics | pca | [25] |
| | | FactoMineR | PCA | [20] |
| | | pcaMethods | pca | [17] |
| sPCA | Sparse PCA | PMA | SPC | [49] |
| | | mixOmics | spca | [25] |
| nsPCA | Non-Negative and Sparse PCA | nsprcomp | nsprcomp | [16] |
| NIPALS PCA | Nonlinear iterative partial least squares analysis PCA | ade4 | nipals | [41] |
| | | pcaMethods | nipalsPca | [17] |
| | | mixOmics | nipals | [25] |
| ICA | Independent Component Analysis | fastICA | fastICA | [116] |
| IPCA | Independent PCA | mixOmics | ipca | [25] |
| sIPCA | Sparse Independent PCA | | sipca | |
| kPCA | Kernel PCA | kernlab | kpca | [19] |
| | | dimRed | kPCA | [129] |
| | | mixKernel | kernel.pca | [18] |
| pPCA | Probabilistic PCA | pcaMethods | ppca | [17] |
| bPCA | Bayesian PCA | | bpca | |
| nlpca | Neural Network-based Non-linear PCA | | nlpca | |
| LLE | Locally Linear Embedding | RDRToolbox | LLE | [26] |
| HLLE | Hessian Locally Linear Embedding | dimRed | HLLE | [129] |
| DRR | Dimension Reduction via Regression | DRR | drr | [130] |
| | | dimRed | DRR | [129] |
| MDS | Multidimensional Scaling | base R | cmdscale | [12] |
| | | dimRed | MDS | [129] |
| wMDS | Weighted Multidimensional Scaling | vegan | wcmdscale | [74] |
| nMDS | Non-metric Multidimensional Scaling | MASS | isoMDS | [38] |
| | | vegan | monoMDS, metaMDS | [74] |
| | | dimRed | nMDS | [129] |
| Isomap | Isometric feature mapping ordination | vegan | isomap | [74] |
| | | dimRed | Isomap | [129] |
| | | RDRToolbox | Isomap | [26] |
| CA | Correspondence Analysis | vegan | cca | [74] |
| | | ade4 | dudi.coa | [41] |
| | | FactoMineR | CA | [20] |
| | | ca | ca | [39] |
| MCA | Multiple Correspondence Analysis | MASS | mca | [38] |
| | | FactoMineR | MCA | [20] |
| | | ca | mjca | [39] |
| NCA | Nonsymmetric Correspondence Analysis | ade4 | dudi.nsc | [41] |
| NMF | Non-negative Matrix Factorisation | NMF | nmf | [35] |
| bNMF | Bayesian Non-negative Matrix Factorisation | CoGAPS | GWCoGAPS | [36] |
| Diffusion maps | | diffusionMap | diffuse | [131] |
| | | dimRed | DiffusionMaps | [129] |
| Force directed methods | | igraph | layout_with_* | [132] |
| | | dimRed | FruchtermanReingold | [129] |

**Table 2**. Correlation-based models for single omic datasets that can be implemented in R packages. The short name, full name, R package, function(s) and references per methods are shown. Isomap methods apply non-linear transformations.

There are numerous extensions of classic PCA [12]. PCA of genetic data can be applied using tools like SmartPCA in Eigensoft [13] or Plink [14] that examine correlations across allelic states rather than as a distance matrix. Sparse PCA (sPCA) adjusts the loading vectors to exclude uninformative features using least absolute shrinkage and selection operator (LASSO) penalisation on the SVD of the covariance matrix so that computation is tractable with fewer PCs [15]. Non-negative, sparse PCA (nsPCA) maximises the covariance per PC or across all PCs and can be used via the R package nsprcomp [16] (Table 2). PCA may struggle if $p$ is large or if the PCs are numerous. Probabilistic PCA (pPCA) can be effective for very complex data and can be implemented via the R package pcaMethods [17], as has PCA, Bayesian PCA (bPCA) (Table 2).

Kernel PCA (kPCA) transforms data in a nonlinear manner into a higher dimensional space to determine if groups or outliers are evident. PCA is implemented in this higher-dimensional space with LASSO regularisation and a proximal gradient descent solution for computational efficiency [18]. KPCA can be explored with the R packages mixKernal [18] and kernlab [19], and PCA with FactoMineR [20] (Table 2). Another nonlinear PCA method is the neural network-based nonlinear PCA (nlPCA) that can be run with R package pcaMethods [17] (Table 2).

Independent component analysis (ICA) assumes that the input data is a mixture to be deciphered and so is similar to PCA. However, ICA seeks to find information on the independent components, rather than on the maximum variance. In this way, ICA treats each feature equally, unlike PCA where major PCs are the focus. Consequently, ICA may be more effective if the aim is to discriminate between samples [21], and it has found more biologically meaningful insights than PCA in human gene expression data [22-24]. Independent PCA (IPCA) applies PCA first to determine the loading vectors. It subsequently applies ICA (using fastICA) to obtain independent loading vectors [25]. IPCA can be effective if the data is not normally distributed. An extension of this is sparse IPCA (sIPCA) that applies the L1 penalisation of the covariance matrix containing sparse data as above, and is available in the mixOmics R package [25] (Table 2).

## 1.4 Multi-dimensional scaling (MDS)

Multi-dimensional scaling (MDS) is a group of methods using eigendecomposition to transform data to a new coordinate system with fewer features, while simultaneously minimising the error between these new coordinates compared to the original ones. This attempts to retain the original distances and coordinates as much as possible, contrasting with PCA for which entirely new axes are created such that the new sample coordinates (eigenvectors) are typically very different and thus difficult to relate to the original features. Consequently, MDS is limited by not minimising the feature numbers as much as PCA. MDS uses Euclidean distances to perform linear transformations. Principal coordinate analysis (PCoA) is the classical form of MDS that performs SVD. Non-metric MDS (nMDS) is an extension of MDS that replaces eigenvector decomposition with optimisation methods via a stress function. Another extension of MDS is isometric feature mapping ordination that can be implemented via Isomap, which replaces the Euclidean distances of MDS with geodesic ones [27-28]. Isomap performs nonlinear data scaling based on the shortest path between a pair of nodes representing the samples based on their input data.

## 1.5 Non-negative matrix factorisation (NMF)

Non-negative matrix factorisation (NMF) applies an unsupervised approach to finding signatures across features explaining sample traits. Omics data is suited to NMF because it is generally non-negative. NMF examines a matrix $X$ of samples across their features to produce a series of additive latent variables iteratively [28]. Unlike PCA, NMF treats all features equally and thus may distinguish between samples more effectively than PCA [21]. NMF was adopted from signal processing and is aimed at clustering samples based on minimising the sum of squares errors (SSEs) [29]. NMF factorises an input matrix $X_{n \times p}$ into two non-negative matrices $W_{n \times k}$ and $H_{k \times p}$ as $X_{n \times p} \sim W_{n \times k} H_{k \times p}$ where $W$ has the basis vectors (akin to loadings) for the samples and $H$ has the coefficient vectors (akin to components) for the features. Here, $k$ is called the rank and represents the level of factorisation (components) present such that $K<<P$. $W$ and $H$ are non-negative and can be combined linearly to make $V_{n \times p} = W_{n \times k} H_{k \times p}$, an estimate of $X_{n \times p}$ [28].

$W_{n \times k}$ and $H_{k \times p}$ are gradually optimised based on gradient descent using multiplicative update rules to minimise the error between $V_{n \times p}$ and $X_{n \times p}$: typically, this error is inversely correlated with *K* [28].

NMF is typically run multiple times with a wide range of K values to select the optimal one, which is important because of the non-convex nature of most omic data. This process known as parameter tuning or regularisation. This best *K* can be challenging to objectively determine if differing optimal K values are found each time: this may depend on the initial seeds and the data's heterogeneity [30]. Ideally, 50-200 NMF runs are required for each value of *K* to provide stability in the sample and feature allocations [31]: this may be computationally time-consuming for large datasets. In each run, a consensus matrix $C_{n \times n}$ from the *n* samples is created such that each pair of samples gets a value of one if they are in the same cluster, and a value of zero if not. A cophenetic correlation coefficient (*ρ*) is calculated from this consensus matrix for each run where *0≤ρ≤1*. *K* can be selected as the value maximising the cophenetic correlation coefficient (*ρ*) before *ρ* decreases sharply [31], though methods of error calculation vary. Typically, NMF allocates samples to the *K* ranks based on the features allocated to the top ranks among the *K*, and so may be less effective if there are many orthogonal features [31]. Bayesian approaches to optimisation can be more effective [32-33], though not always [34]. NMF can be implemented with the R packages NMF [35] or CoGAPS [36] (Table 2).

## 1.6 Correspondence Analysis (CA)

Correspondence Analysis (CA) is a correlation- and projection-based method examines ordinal data, such as a matrix containing non-continuous binary omic data, and creates orthogonal components for each sample. CA uses measures of inertia and co-inertia. In simple terms, inertia refers to the sums of squares (SS) of distances between a set of points and a reference point, or the SS of matrix elements in a zero-centred matrix [11]. Co-inertia extends this by denoting the SS of covariances between two mean-centred datasets [37], thus indicating their pairwise association. CA zero-centres and scales the data in a contingency table manner. It then performs SVD on this transformed data for visualisation. CA becomes multiple CA (MCA) if more than two categorical variables are assessed. CA and MCA can be applied with the R packages MASS [38], ca [39] and FactoMineR [20] (Table 2). Multiple CA (MSC) and discriminant CA can be effective with qualitative data. Nonsymmetric CA (NCA) is a variant of CA to the symmetrical relationship of two ordinal variables [40] and can be used in R package ade4 [41] (Table 2). CA is limited by not tolerating continuous data well, may be affected by data transformation, and assumes independence among the features.

## 2. Correlation-based dimension reduction methods for pairs of omics datasets

Pairs of omic datasets must be transformed and normalised independently in an appropriate manner to ensure comparability because the shared components will reflect both datasets. Differing numbers of samples and features per omic dataset should be tolerated. The two datasets can be weighted according to a number of schemes: equally, by importance, data quality, or proportionally to the number of features [11]. Dimension reduction methods may also examine three or more omic datasets – the latter are K-table methods such that *K* refers to the number of omic datasets [42]. Thus, each of *K* datasets would be scaled by a global score matrix containing the components and each would have *K* different loading vectors. Although dimension reduction and analysis of each omic dataset alone would extract more information about these datasets individually, early integration of omic datasets is essential to extract information across these diverse inputs.

## 2.1 Consensus PCA (cPCA)

PCA implemented most typically via SVD, but also may be accomplished using the iterative algorithm Nonlinear Iterative Partial Least Squares (NIPALS) that determines the next PC per iteration (NIPALS PCA) across block and global loading scores based on the root mean squared error (RMSE). Consensus PCA (cPCA) is a projection-based unsupervised method that uses the NIPALS algorithm where PCA is conducted on normalised combined data to get the relative weights of the different omic datasets (called blocks here) [43]. cPCA is a type of multi-block method in which each set of omic data is a normalised block in the combined model [44]. This iterates by first regressing the omic datasets on a global score vector to create global loadings reflecting the weight of each block. Second, these weights are normalised and are used to generate new block scores. These are regressed on the global score vector to create a normalised vector of weights, which is used to create the update block score values for the next iteration. Such block scaling factors can be effective when the number of features per omic type varies considerably, so these weights might be inversely proportional to the number of features per block. This links to multiway methods where multiple blocks are stacked to make an N-dimensional dataset. An application of this compares between omics data types with multilinear Partial Least Squares (PLS) [45]. cPCA can be applied with the mogsa R package [46] and NIPALS PCA is in numerous packages, including pcaMethods [17] (Table 3). The number of iterations, impact of subsampling and structure in the data can all affect cPCA results.

| Name | Full method name | Package | function(s) | Reference |
|---|---|---|---|---|
| cPCA | Consensus PCA | mogsa | mbpca | [46] |
| CCA | Canonical Correlation Analysis (CCA) | CCA | cc | [155] |
| | | vegan | CCorA | [74] |
| | | PMA | CCA | [49] |
| rCCA | Regularised CCA | CCA | rcc | [155] |
| sCCA | Sparse CCA | SmCCNet | getRobustPseudoWeights | [52] |
| smCCA | Sparse Multiple CCA | PMA | MultiCCA | [49] |
| ssCCA | Sparse Supervised CCA | SmCCNet | getRobustPseudoWeights | [52] |
| CIA | Co-Inertia Analysis | made4 | cia | [57] |
| | | | cointeria | |
| STATIS | Structuration des Tableaux a Trois Indices de la Statistique | ade4 | statis | [41] |
| PTA | Partial Triadic Analysis | | pta | |
| statico | STATIS and CIA | | statico | |
| MANCIE | Matrix analysis & normalisation by concordant information enhancement | MANCIE | mancie | [62] |

**Table 3**. Unsupervised correlation-based models for omic dataset pairs that can be implemented in R packages. The short name, full name, R package, function(s) and references per methods are shown.

## 2.2 Canonical Correlation Analysis (CCA)

CCA is a projection-based and correlation-based unsupervised extension of CA [47]: here, CCA refers to Canonical Correlation Analysis (rather than Canonical Correspondence Analysis) [11]. CCA linearly transforms a pair of omic datasets to maximise correlations across their features based on a pair of loading vectors (also called canonical variates or dataset-specific weights). CCA is optimised based on the canonical correlation from this pair of loading vectors. CCA is effective if the number of samples exceeds the sum of the total number of features across omic types, or if the number of features has been reduced [48]. Regularised CCA (rCCA) can tolerate total numbers of features well in excess of the number of samples, and thus is more suited to omic data. It achieves this by making the omic data matrices invertible through the addition of ridge penalties to their diagonals. The optimal ridge penalties can be obtained using cross validation if the number of features per dataset < 5,000, or using shrinkage where they are independently estimated and so their correlation might be affected. Consequently, the numbers of features, parameter selection and potential model overfitting are possible limitations of CCA and rCCA. CCA and rCCA can be implemented via the mixOmics R package [28] (Table 3).

There are many other extensions of CCA, including penalised CCA, sparse CCA (sCCA), CCA-l1, sparse CCA with an elastic net penalty (CCA-EN), CCA-group sparse [49] and a supervised version of sparse CCA, ssCCA [50]. sCCA is an unsupervised approach that regularises the data through the addition of a ridge penalty and is similar to tensor decomposition [65]. Perturbation clustering for data integration and disease subtyping (PINS) can extend sCCA by using clustering samples within and across different omic datasets where these clusters are robust to noise [51]. ssCCA and sparse multiple CCA (smCCA) are supervised methods, where smCCA reweights the features to maximise the correlation between the quantitative sample metadata with each group's features [52]. CCA has been extended by examining sample metadata classifications in sCCA using networks (SmCCNet) [52]. SCCA, ssCCA and smCCA can be used via the R packages PMA (penalised multivariate analysis) [49] and SmCCNet [52] – the latter extends PMA to handle more than two omic datasets (Table 3).

## 2.3 Co-inertia analysis (CIA)

Co-inertia analysis (CIA) is a correlation- and projection-based unsupervised method that seeks to maximise the covariance between a pair of transformed datasets that can be combined linearly [53-54]. CIA is a generalisation of CCA and PLS [55] but may be limited by the need to have nonzero loading vectors [56]. CIA does not have a feature selection step, unlike sparse CCA, which can remove uninformative features. CIA is available in the R package made4 [57] (Table 3). The ade4 R package offers a range of additional less commonly-used tools, including Structuration des Tableaux a Trois Indices de la Statistique (STATIS) [58], partial triadic analysis (PTA) [59], and statico that can find relationships between two pairs of K-tables using STATIS and CIA [8,41] (Table 3).

Matrix analysis and normalization by concordant information enhancement (MANCIE) is related to CIA. This method achieves data normalisation and fusion of dissimilar omic datasets by linking a pair of omic datasets prior to the main analysis based on the feature patterns [60] (Table 3). If the features are not directly comparable between the primary and secondary datasets, it summarises the secondary data so that the features can be combined into a single primary-secondary omic dataset. If the primary and secondary features of sample $i$ are $p_i$ and $q_i$, the combined dataset features of sample $i$ then are the 1st PC from PCA if $cor(p_i,q_i) > t_U$, or equal to $p'_i$ if $cor(p_i,q_i) \leq t_L$, or equal to $p'_i + q'_i.cor(p_i,q_i)$ if $t_L < cor(p_i,q_i) \leq t_U$. The upper ($t_U$) and lower ($t_L$) thresholds can be determined empirically, and $p'_i$ and $q'_i$ are scaled such that their standard deviations equal one [60]. Like other methods, MANCIE may be affected by data transformation, parameter choice and is prone to overfitting.

## 2.4 Partial least squares (PLS) discriminant analysis (DA)

PLS is similar to PCA because it applies SVD and iteratively maximises the covariance across the features [61]. PLS is a correlation- and projection-based method that can be applied to a single dataset, or to pairs of them by concatenating data together. PLS may be limited by feature dependence, model overfitting and optimising the number of ascertained components. Sparse PLS (sPLS) can also be useful when a large number of features is present as it can exclude uninformative ones using LASSO penalization on the SVD of the omics dataset pair's covariance matrices [15]. sPLS and CCA-EN typically have comparable predictive performances that are superior to CIA [25]. Group PLS (gPLS) is an extension that applies pre-defined groups of features to PLS analyses [62].

PLS-discriminant analysis (PLS-DA) is a supervised method that incorporates group memberships into PLS [63]. These memberships could include genomic allelic states or protein-protein binding categories instead of (or in addition to) traditional group labels. PLS-DA is akin to supervised PCA because it tries to combine correlated features to discriminate between known groups [64]. PCA maximises the covariance per PC, whereas PLS-DA maximises the covariance based on the original classification [64]. Intuitively, this means PCA and PLS-DA produce similar results if the 1st (and 2nd, etc) PC(s) include(s) features that discriminate between groups, whereas they differ if the groups are more similar with only minor differences detected at higher PCs by PCA. PLS-DA aims to determine the pair of loading vectors that maximise the covariance between the two omics datasets' values, $X$ and $Y$. So, for matrices $X_{n \times p}$ and $Y_{n \times q}$, PLS-DA maximises the correlation between the loading vectors associated with $p$ and $q$ and iterates through the PCs. These loading vectors (projections) in PLS-DA reflect information to discriminate group status, thus informing inter-group differences, but less so within-group differences [65]. PLS-DA uses a LASSO-based penalty for selection of the loading vector weights. It can be implemented in regression mode where X explains Y or in a canonical mode, where X and Y can explain each other (assuming X and Y are symmetrical to allow switching). PLS-DA and a sparse implementation for large datasets are available through the caret [66] and mixOmics [28] R packages (Table 4).

| Name | Full method name | Package | function(s) | Reference |
|---|---|---|---|---|
| PLS | Partial Least Squares (PLS) | mixOmics | pls | [25] |
| sPLS | Sparse PLS | mixOmics | spls | [25] |
| PLS-DA | PDS Discriminant Analysis | mixOmics | plsda | [25] |
| sPLS-DA | Sparse PLS Discriminant Analysis | | splsda | |
| | | caret | | [67] |
| mbPLS | Multi-block PLS | ade4 | mbpls | [41] |
| mgPLS | Multi-Group PLS | mixOmics | mint.block.* | [25] |
| GPA | Generalised Procrutes Analysis | vegan | procrutes | [74] |
| | | FactoMineR | GPA | [20] |

**Table 4**. Supervised correlation-based models for omic dataset pairs that can be implemented in R packages. The short name, full name, R package, function(s) and references per methods are shown.

To explore how omic dataset pairs can be examined together, we can use PLS-DA in the mixOmics R package to quantify how gene expression and protein-binding rates relate to each other in 100 samples from four groups. PLS-DA identifies the features most strongly differentiating these groups using small initial weights (set to 0.1) while ignoring within-group variation. The features most strongly correlated between omic types can be identified for further analysis (Fig 2A). The top 25 differentiating features can be used to separate groups: here this works well for groups A and B, but less well for groups C and D (Fig 2B). If we expand the number of omic types from two to four by including data on DNA genotypes and interferon-gamma (INFg) expression, we can explore correlations between the highly correlated features (Fig 2C). Similarly, the groups can be classified using a select number of features (Fig 2D). This illustrates how the features differentiating samples for different numbers of input omic types can vary, and therefore the weighting of these features is crucial.

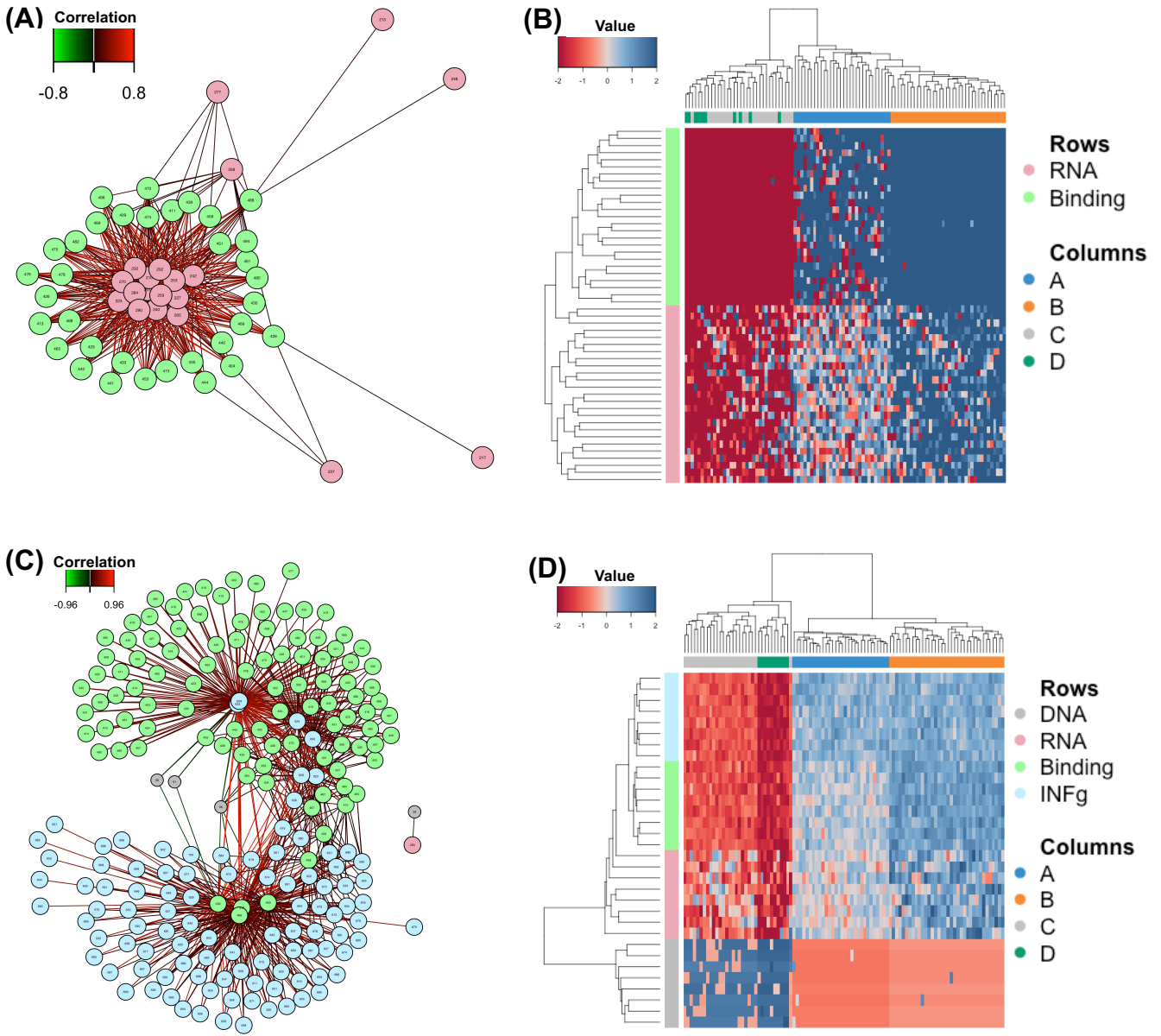

**Fig 2**. Differentiation of 100 samples using combined analyses of RNA expression and protein-binding data with PLS-DA. (A) A network showing correlations between RNA expression (red) and protein-binding (green) where *r < -0.7* or *r > 0.7*: positive correlations are shaded green and negative ones are shaded red. (B) A heatmap of the normalised RNA expression and protein-binding rates of the samples corresponding to groups A (n=30, blue), B (n=26, orange), C (n=24, grey) and D (n=20, green) where red indicates a lower value, white an intermediate one, and blue higher values; the dendrograms indicate the relative similarity of the features (y-axis) and samples (x-axis); the 25 most differentiating features for each omic type are visualised. (C) A network like (A) showing correlations where *r < -0.9* or *r > 0.9* between DNA genotypes (black), RNA expression (red), protein-binding (green) and INFg expression (blue). (D) A heatmap like (B) of the normalised DNA genotypes, RNA expression, protein-binding rates and INFg expression of the samples; the eight most differentiating features for each omic type are visualised.

Multi-group PLS (mgPLS) extends PLS by using global loading vectors for the omic datasets [67]. Approaches integrating distinct omic datatypes include weighted-averaging PLS (WA-PLS) is a regression-based method, which uses only the first component from PLS, and MINT [68], which removes unwanted variation stemming from batch effects using PLS-DA and mgPLS. MINT is in the mixOmics R package and applies multi-group and generalised PLS-related methods, including sparse versions [68] (Table 4). However, PLS-DA is prone to overfitting and finding unreliable features distinguishing groups during supervised clustering [69], especially if there are fewer samples relative to many features [70].

## 2.5 Generalised Procrustes analysis (GPA)

Generalised Procrustes analysis (GPA) is a projection-based supervised method that linearly transforms a primary data matrix to maximise its similarity to a secondary matrix by minimising the SS of differences [71]. This least-squares-based exploratory method achieves this by first mean-centring the datasets, it then scales (root mean square distance) them to make the values comparable and finally rotating them via SVD to minimise the SS of differences between the corresponding data points. GPA may be limited by the effects of outliers, data structure and datasets with high numbers of uncorrelated features. GPA can be implemented via the R packages vegan [74] and FactoMineR [20] (Table 4).

## 3. Correlation-based models for more than two omic datasets

The methods below were designed for more than two omic datasets, however, most could also be applied to pairs of datasets or paired ones.

## 3.1 Multiple factor analysis (MFA)

Multiple factor analysis (MFA) can be used to examine multiple omic datasets and has multiple implementations [72]. Most commonly, generalised PCA is applied to each omic dataset individually, and these datasets are normalised by dividing by their first singular values [73]. These datasets are concatenated and generalised PCA is applied again. MFA, dual MFA and hierarchical MFA can be applied via the R packages FactoMineR [20] and ade4 [44] (Table 5).

## 3.2 Multiple CIA (mCIA)

Multiple co-inertia analysis (mCIA) extends CIA from assessing a pair of datasets to examining three or more datasets [74]. mCIA transforms the omic datasets using loading vectors to maximise the sum of their covariances and makes a new synthetic centre of all data [75]. This can be iterated to get the global scores and loadings for the first dimension, before doing this for the second dimension, and so on. It is similar to cPCA because it maximises the covariance between the eigenvectors, and produces similar results to generalised CCA [11]. MCIA is a correlation- and projection method that is available in the R package omicade4 [75] (Table 5).

MCIA may work less well with sparse data [11]. To address this, sparse mCIA (smCIA) selects features and estimates the loading vectors simultaneously to address this sparsity issue, and structured smCIA (ssmCIA) extends this by including a penalty to include biological information more effectively [76]. Similarly, penalised CIA (pCIA) imposes a sparsity penalty [56]. To illustrate how information can be aggregated across omic types per sample, we can use an example of data for DNA genotypes, gene expression, protein-binding rates and interferon-gamma levels for 100 samples from four groups. Here, mCIA can quantify the correlations per sample between omic types, the extent to which each feature is associated with each component, and the contribution of each omic type to each component (Fig 3).

## 3.3 Joint and individual variation explained (JIVE)

Joint and individual variation explained (JIVE) extends PCA by partitioning the total variation found into variation unique to each omic data type, variation shared across data types, and residual noise [61]. It assumes the observed data can be explained by a linear combination of unique, shared and residual variation where each variation type is uncorrelated with the others. It applies a linear model, where $w_{nk}$ is the sample's score, $h_{kp}$ is the estimated transformed features, $\varepsilon_{nk}$ is the error term, so for feature $p$ in sample $n$ as: $x_{np} = \sum_{k=1}^{K} w_{nk} h_{kp} + \sum_{k=1}^{K'} w'_{nk} h'_{kp'} + \varepsilon_{nk}$, where $w'_{nk}$ and $h'_{kp}$ are technology-specific vectors whose dimensions span $n$ samples and $K'$ components for $W$, and $K'$ components and $p'$ features for $H$, such that $p' < p$ and $K' < K$ and $W'$ and $H'$ are orthogonal to $W$ and $H$. JIVE iteratively estimates the unique and joint variation by SVD focusing on one term per iteration so the next one has maximum orthogonality. JIVE may be limited by the need to select components and the loss of information after this, as well as the effects of data transformation. JIVE is a correlation- and projection-based approach that can be applied using the r.JIVE R package [77]. JIVE and MFA can examine concatenated omic data from different types for the same samples (Table 5).

| Name | Full method name | Package | function(s) | Reference |
|---|---|---|---|---|
| MFA | Multiple Factor Analysis | FactoMineR | MFA | [20] |
| DMFA | Dual Multiple Factor Analysis | | | |
| HMFA | Hierarchical Multiple Factor Analysis | | | |
| MFA | Multiple Factor Analysis | ade4 | mfa | [41] |
| mCIA | Multiple Co-Inertia Analysis | omicade4 | mcia | [75] |
| JIVE | Joint & Individual Variation Explained | r.jive | jive | [79] |
| AJIVE | Angle based JIVE | idc9 | ajive | [80] |
| rGCCA | Regularised Generalised CCA | RGCCA | rgcca | [82] |
| | | mixOmics | wrapper.rgcca | [25] |
| sGCCA | Sparse Generalised CCA | RGCCA | sgcca | [82] |
| | | mixOmics | wrapper.sgcca | [25] |
| ssGCCA | Supervised sparse Generalised CCA | | DIABLO | [86] |
| O2PLS | Two-way Orthogonal PLS | OmicsPLS | o2m | [108] |
| GO2PLS | Group Sparse Two-way Orthogonal PLS | | | [109] |
| PO2PLS | Probabilistic Two-way Orthogonal PLS | | | |
| TCA | Tensor component analysis | tensorBSS | tPCA | [102] |
| | | rTensor | cp_decomp | [103] |
| | | tensorr | dtensor | [105] |
| | | ThreeWay | CP | [101] |
| | | ThreeWay | T3 | |
| | | PTAk | PCAn | [100] |
| | | PTAk | CANDPARA | |
| | | SDA4D | RunSDA4D | [108] |

**Table 5**. Unsupervised correlation-based models for more than two omic datasets that can be implemented in R packages. The short name, full name, R package, function(s) and references per methods are shown. OmicsPLS and idc9 should be installed from Github with devtools.

JIVE has been extended as joint and individual clustering analysis (JIC) that estimates the unique and joint variation simultaneously in a manner similar to combined PCA and k-means clustering [78]. It has also been extended as angle-based JIVE (AJIVE), which finds joint modes of variation common to all omics data types, and variability specific to each omic type [79]. AJIVE can be applied with R package idc9 [80]. Using the same dataset above used for mCIA, AJIVE partitions the variation per omic type: 2.5-3.5% of the total variation was allocated to joint variation for DNA genotypes, RNA expression and protein-binding, whereas 11% for INFg levels (Fig 3). The individual variation allocated to each omic type was zero for DNA and INFg variability, but high at 83% for RNA and 78% protein-binding. The residual variation was mainly associated with the DNA genotyping (96%) and INFg levels (89%), but less so for RNA expression (14%) and protein-binding rates (19%). This highlights the differing levels of correlated features across omic datasets, and the diverse levels in which features and samples may be positively, negatively or un-correlated across omic types.

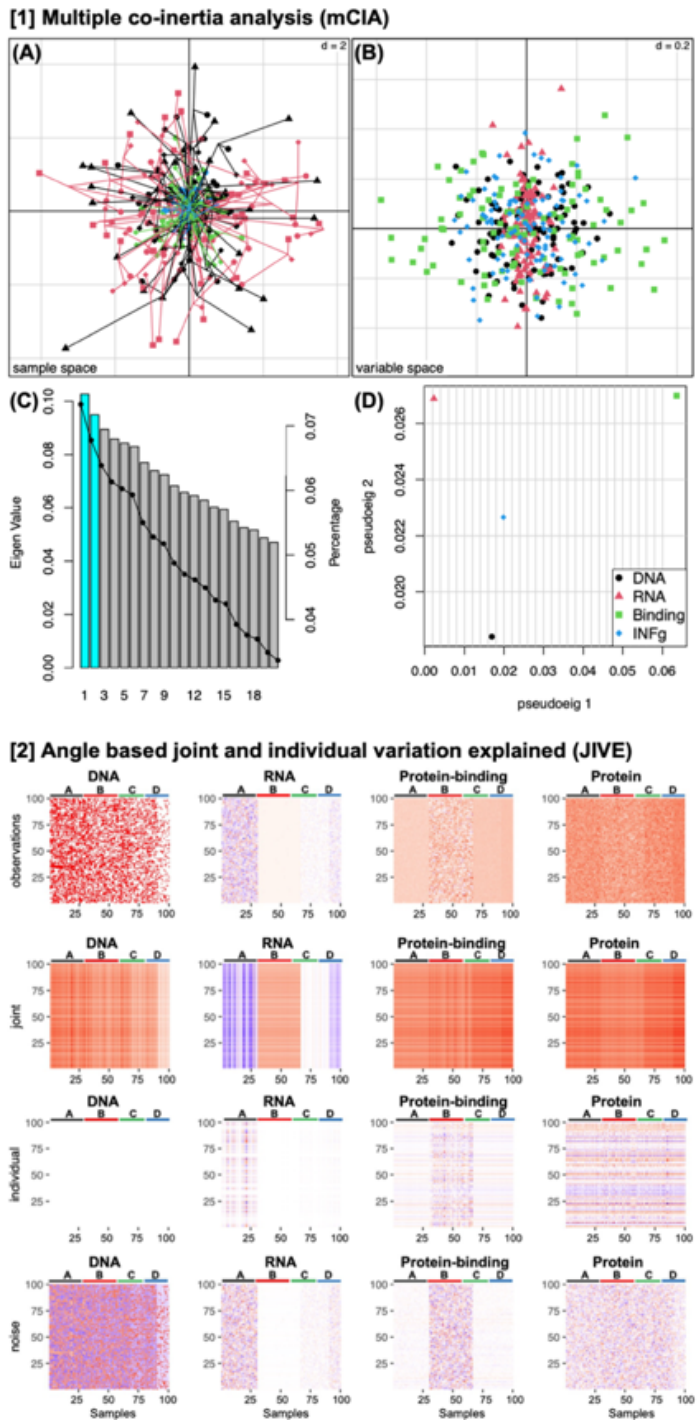

**Fig 3**. Analyses of 100 samples corresponding to groups A (n=30, black), B (n=36, red), C (n=24, green) and D (n=10, blue) assessed integrating four omic datasets: DNA, RNA, protein-binding and interferon-gamma expression using [1] mCIA (top) and [2] AJIVE (bottom). This data has slight differences for each omic type that are correlated across omic types. [1] (A) A mCIA plot showing the sample space for these 100 samples showing their omic types: DNA (circle), RNA (triangle), protein-binding (square) and interferon-gamma (INFg) rates such that a shorter edge connecting a pair of samples from different omic datasets reflects a more positive association between them. The RNA levels show higher heterogeneity. [1] (B) A mCIA plot showing the feature space in which features more strongly associated with a group are shown on more positive coordinates, whereas those with a lower association are on negative coordinates. Here, the protein-binding activity shows high variability. [1] (C) The sorted eigenvalues for each eigenvector where the cyan colour denotes the plotted eigenvalues (1 and 2) and the black dots represent the fraction of variance explained by each eigenvalue. [1] (D) The pseudo-eigenvalues space shows the extent to which each omic type contributes to eigenvalues 1 and 2: DNA diversity affects eigenvalue 1 more, whereas eigenvalue 2 is more associated with RNA activity and INFg levels. [2] The AJIVE results show the underlying data (1st panel), which ranges from present (red) and absent (white) for the DNA genotypes, and low (blue), white (intermediate) or high (red) for the RNA, protein-binding and protein levels. In the 2nd panel, the joint variation shared across omic types that measurements were most variable for the RNA activity. In the 3rd panel, we see the variation associated with each individual omic type, which was higher for RNA activity for group A and protein-binding for group B. In the 4th panel, noise was higher for the DNA genotypes and protein levels, the RNA activity for group A, and the protein-binding for group B, but reduced for the other datasets.

## 3.4 Canonical Correlation Analysis (CCA) extensions to multiple datasets

Like sCCA, generalised CCA (gCCA) extends CCA to more than a pair of omic datasets and can include a feature selection step for large sparse datasets [1]. gCCA is somewhat related to MFA. Multiple CCA (mCCA) can deal with more than two omic datasets and is an extension of PLS that maximises the correlation between the omic datasets' loading vectors [50]. Probabilistic CCA (pCCA) can be applied to learn shared and unique properties across omic datasets, followed by factor analysis of the residuals, also called Multiset Correlation and Factor Analysis (MCFA) [81]. Regularized generalized CCA (rGCCA) has also been developed as an analysis framework [82]. Sparse generalised canonical correlation analysis (sGCCA) is an extension of rgCCA that can be implemented via the RGCCA R package [83], as can sGCCA (Table 5). Tensor CCA (tCCA) attempts to maximise the canonical correlation of multiple data sources [84], and was extended as tensor sparse CCA (TSCCA), discussed later [85]. SGCCA has an extension (DIABLO in the mixOmics R package) that can handle supervised tests [86] (Table 5). For this, pre-processed and normalised omic datasets, a known numbers of components to assess, and a design matrix specifying the underlying structure of interest are required. Alternatively, the design matrix can be inferred from PLS, modelling the pairwise associations between omic datasets [87]. This supervised sGCCA approach identifies correlated features across the omic data sources to select appropriate components per feature set by maximising the correlation across features. It identifies these components within and between omic datasets iteratively using L1 penalisation.

## 3.5 Tensor component analysis (TCA)

Tensor component analysis (TCA) is a broad set of projection-based methods and is sometimes called tensor decomposition, tensor rank decomposition, canonical decomposition (CANDECOMP) [87] and parallel factor analysis (PARAFAC) [88], CANDECOMP/PARAFAC (CP), Tucker3, and other names depending on the precise implementation approach [89]. TCA is based on the principle that a sample's omic data sources are structured and correlated in some manner: for instance, that a gene's expression rate is associated with its protein level, or a gene's expression rate may be similar across tissues. TCA uses tensors, which have an order (or dimension number) $n$ such that if $n=0$ the tensor has zero dimensions and is single value; if $n=1$ the tensor has one dimension and is an array; if $n=2$ the tensor has two dimensions and is a matrix; and if $n=3$ the tensor has three dimensions, and so on. Consequently, tensor-based methods have the advantage of manipulations of tensors as vectors or multi-dimensional matrices (e.g., containing data on $n$ samples across $p$ genes across $j$ omic datasets) [90]. A disadvantage of TCA is that it is computationally intensive, although this can be mitigated through the use of multi-view matrices for feature extraction [91].

Many TCA approaches are based on Tucker decomposition, which is an extension of SVD to higher-dimensional data. Here, we focus on third-order tensors as an example. Tucker decomposition works by decomposing (for example) omic data represented by the tensor $A_{n \times p \times t}$ ($A \in R^{n \times p \times t}$) for $p$ genes across $n$ samples from $t$ timepoints into a set of matrices and a small core tensor that relates $t$ (also known as modes) to one another [92]. If we have three input omic datasets ($X_{n \times p}, Y_{n \times p}, Z_{n \times p}$), this can be decomposed into $n \times d$ matrices where the core tensor is $T_{d \times d \times d}$ and $d \leq n$ such that $A_{n \times p \times t} = T_{d \times d \times d} U_{n \times d} V_{p \times d} W_{t \times d}$ and typically $d << n$ for computational efficiency (Figure 3). $U$, $V$ and $W$ are factor matrices. The product $T_{d \times d \times d} U_{n \times d} V_{p \times d} W_{t \times d}$ approximates $A_{n \times p \times t}$ as $\hat{A}_{n \times p \times t}$ where the algorithms minimise the difference between $A$ and $\hat{A}$, usually using an alternating least squares (ALS) approach. If the features in $X$, $Y$ and $Z$ are actually $p_1$, $p_2$ and $p_3$, then $d$ can vary as $d_1$, $d_2$ and $d_3$: $A_{n \times p \times t} = T_{d1 \times d2 \times d3} U_{n \times d1} V_{p \times d2} W_{t \times d3}$. This is a form of higher-order SVD (HOSVD) is a special form of Tucker decomposition where the core tensor and factor matrices are orthogonal.

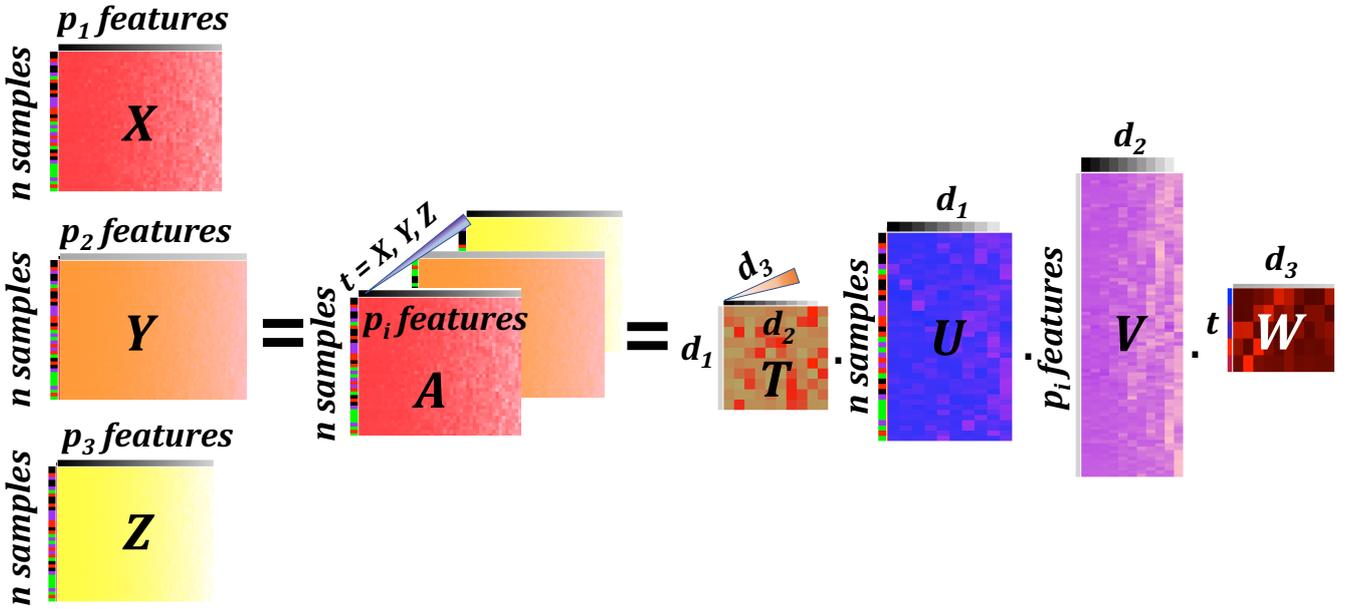

**Fig 3.** Illustration of Tucker decomposition. Three input omic datasets with *n* shared samples ($X_{n \times p1}$ in red, $Y_{n \times p2}$ in orange, $Z_{n \times p3}$ in yellow) with $p_1$, $p_2$ and $p_3$ features each, respectively, can be decomposed as a tensor $A_{n \times pi \times t}$ with *t=3* into core tensor $T_{d1 \times d2 \times d3}$ (brown/red) with components $d_1$, $d_2$ and $d_3$, and three factor matrices ($U_{n \times d1}$, $V_{p \times d2}$, $W_{t \times d3}$).

PARAFAC is similar to Tucker decomposition above in that it generates three two-dimensional matrices as an individual scores matrix $X_{n \times c}$, a scores matrix $Y_{t \times c}$, a sparse gene loadings matrix $Z_{c \times p}$ and an error matrix $\varepsilon_{npt}$ for components *c*. It uses a linear model to decompose each tensor value $A_{npt}$ across all components *C* as: $A_{npt} = \sum_c^C X_{nc} Y_{tc} Z_{cp} + \varepsilon_{npt}$. Multilinear PCA (MPCA) (also known as M-mode PCA) and multilinear ICA (MICA) are similar to this. MPCA creates a matrix for each mode (ie, each omic data type) where each of the column entries are orthogonal to one another (ie, the matrix is orthonormal). The variation retained in MPCA is maximised across the whole tensor (rather than per matrix), iterated using ALS.

TSCCA (tensor sparse CCA) uses Tucker decomposition, and was applied to miRNA-RNA-cancer associations, but the principles are more broadly applicable if the features are matched [86]. It decomposes a tensor of correlation values $A_{n \times pi \times q}$ ($A \in R^{n \times pi \times q}$) for *n* samples across $p_i$ genes and *q* other features (e.g. expression levels). It counts the numbers of nonzero elements observed to ensure sparsity. The average Pearson correlation across each *n*, *p* and *q* combination is used as a proxy for modularity to form the tensor.

2D-PCA and generalised 2D-PCA (G2D-PCA) seek to extend PCA by directly determining the eigenvectors of the covariance matrix with matrix-vector conversion [93-94]. This has been developed in to generalised N-dimensional PCA (GND-PCA) that encodes the data as an *n*-order tensor (for *n≥3*) containing the layers of omic data and uses HOSVD to get information on each omic subspace [95].

One approach applying this is sparse decomposition of arrays (SDA, also called sparse tensor factorisation or sparse matrix factorisation). This decomposes omic data $A_{n \times p \times t}$ (as above) for $p_i$ genes across *n* samples from *t* timepoints (modes) into *d* latent components as a pair of *n×d* and *d×p_i* sparse matrices such that $d \ll p_i$ [96]. These sparse matrices can be initially determined with fastICA (R package fastICA) [97] fitted using Variational Bayes modelling (Table 5). Here, *t* could equally represent gene expression over time for longitudinal data. SDA has been applied to GWAS where SNP genotypes are analysed in each latent component *d* in *n*×*d* [96]. This flexible approach can handle variation in the numbers of features across *t*.

Another method called TCAM (TCA for multi-way data) uses unsupervised tensor factorisation to generalise SVD for matrices to high-order tensors [98]. TCAM requires prior knowledge of the number k of latent components and can measure longitudinal intra-sample (as well as inter-sample) variation [98]. Unlike TCAM, the truncated SVD method (tSVDM) decomposes a tensor $A_{n \times p \times t}$ into three matrices: one with information about the samples ($U_{n \times n \times t}$), one with the singular values ($S_{n \times p \times t}$), and one with information about the features ($V^T_{p \times p \times n}$). In contrast, TCAM uses the tensor $A_{n \times p \times t}$ as a set of *n* *p*×*t* matrices and makes *q* feature matrices that are similar to PCs, each with loadings (eigenvectors) and scores, which might reduce the number of dimensions from *t* to *q* [98].

Non-negative tensor factorisation (NTF) can apply a range of algorithms to solve each of the *q* sub-problems associated with factorising a tensor. Multi-omics non-negative tensor decomposition for integrative analysis (MONTI) implements NTF through a trilinear model in PARAFAC first, and next applies L1 regularisation to get components associated with sample subsets, where these components are used to make a classification system [99]. MONTI focuses on human transcriptome, methylome and miRNAome data, and so gets the features associated with these components for further pathway-oriented testing.

Manipulations of tensor-valued data can be achieved in various ways with the R packages PTAk (Principal Tensor Analysis on k Modes) [100], ThreeWay [101], tensorBSS [102], rTensor [103], and TensorFlow [104] (Table 5). Additionally, R package tensorr [105] can handle tensors with sparse data, and SDA4D can examine four-dimensional tensors [106] (Table 5).

### 3.6 Orthogonal PLS-DA

Orthogonal PLS-DA is an extension that can handle more than two omic datasets that is most commonly implemented as two-way orthogonal PLS (O2-PLS). This estimates the PLS components based on a fraction of a primary omic dataset explained by the others across all omic data pairs [107]. For each, it allocates the observed variation into three classes: joint variation based on the covariance between omic types, independent variation unique to each type, and residual variation (error). O2-PLS can be implemented via the R package OmicsPLS [108] (Table 5). It can be effective for low-complexity data if there are sufficient PCs explaining a majority of the dataset, which can be explored by plotting the loadings of the covariance per omic dataset pair [108]. In the same package are probabilistic O2PLS (PO2PLS) and group sparse O2PLS (GO2PLS). GO2PLS allows sparsity in the input data and achieves this by removing features approximating zero and can identify joint variation shared between omic types for a subset of all the samples [109] (Table 5). This has been extended further to assess more than two omic data types in OnPLS [110-111] and sparse multi-block (MB) PLS (sMBPLS) [112], where the latter extends PLS of K-tables and is suited to CNV, methylation, expression and miRNA data (Table 5). SMBPLS determines multi-dimensional regulatory modules based on maximising the covariance across omic datasets, using linear models in an iterative fashion for each PLS predictor [113].

### 4. Network-based models for more than two omic datasets

### 4.1 Similarity network fusion (SNF)

Similarity network fusion (SNF) works by creating networks for each omic data type depicting the samples as nodes and creating edges between these nodes based. First, the data is normalised, mean-centred and transformed into similarity graphs based on their relative initial Euclidean distances for each individual omic dataset [114]. Second, global (across all datasets) and local (within each dataset) similarity matrices are created based on these network distances. Samples with the same feature are collapsed into a single node iteratively based on message passing [115]. This nonlinear nearest-neighbour approach iterates until it converges on a single similarity network across all the omic data types. SNF can work well with expression, methylation and miRNA data [115] and is available in R package SNFtool [116] (Table 6). SNF is a correlation- but not projection-based method.

| Name | Full method name | Package | function(s) | Reference |
|---|---|---|---|---|
| SNF | Similarity network fusion | SNFtool | SNF | [116] |
| WGCNA | Weighted correlation network analysis | WGCNA | (various) | [117] |
| netOmics | Integrated network generation | netOmics | (various) | [156] |
| timeOmics | Linear mixed model | timeOmics | (various) | [4] |

**Table 6**. Network-based models for more than two omic datasets that can be implemented in R packages. The short name, full name, R package, function(s) and references per methods are shown.

### 4.2 Weighted correlation network analysis (WGCNA)

Weighted correlation network analysis (WGCNA) examines the association between the features and the traits belonging to the samples as a correlation coefficient. This widely-used method summarises correlated features as single nodes or modules in a network, and examines their relative distances in the network. Typically, the features examined are gene expression levels, though the principles could be applied more broadly. WGCNA can be implemented with the R package WGCNA [11,117] (Table 6) and is a correlation- but not projection-based method.

### 4.3 Longitudinal expression data network-based integration

Network-oriented approaches to understanding time-series expression have been developed based on linear mixed models and network propagation. One that can be applied with R package netOmics first develops a (potentially unique) linear model for each molecule that also eliminates missing timepoints/features [4] (Table 6). Second, it clusters the molecules based on their relative model similarity for each omic type using MB PLS [4]. Third, it constructs data- or knowledge-driven networks for each omic dataset and then integrates these into a single network representation using the random walk with restart (RWR) algorithm in R [118]. This is a correlation- but not projection-based approach that has been extended by the related R package timeOmics to include a final validation step to check cluster robustness [4] (Table 6).

### 4.4 Topological data analysis (TDA)

Topological data analysis (TDA) of omic data networks can assess high-dimensional features, and has been applied to diverse genome [119], transcriptome [120] and protein-protein interaction (PPI) [121-122] datasets. TDA transforms an omic dataset into a simplicial complex to decipher the underlying topological structures, such as the connection patterns among protein modules in a PPI network. Thus, TDA focuses on shapes such that network components can be rotated about their nodes without the loss of information. The complexity of omic data means that computationally efficient TDA approaches are important. One common simplicial complex construction method is a Vietoris-Rips filtration [123] that connects all nodes with a specific distance of one another. TDA can quantify and analyse many network features including the numbers of nodes, edges, indirect connections, triangles, tetrahedra and high-dimensional polytopes [124].

## 4.5 Other multi-omic analysis approaches

There are a variety of Bayesian approaches, including unsupervised ones that factorise data into latent factors such as Multi-Omics Factor Analysis (MOFA) [125] and Multiple Dataset Integration (MDI) [126]. MOFA uses this latent space to infer missing values [124], but this might affect the independence of the observed feature values. Methods based on Bayesian networks iteratively change the networks constructed from the variables to optimise the fit. An example of this for continuous data is a conditional Gaussian Bayesian network (CGBN) that add omic layers iteratively and tests the fit with Bayes Factors. Lastly, Bayesian consensus clustering (BCC) focuses on differences between clusters within (source-specific) and between (common) omic data types using hierarchical Dirichlet mixture model and Gibbs sampling estimation [127]. It assigns the samples to clusters using Gibbs sampling to estimate the parameters, allocating the samples to the same ones across omic data types where possible. BCC can be applied with R package bayesCC via devtools [127]. BCC and MF methods perform well at determining the correct number of clusters in simulated data [2]. Other approaches use network-based methods to look for patterns across pathways, biological databases to map transcripts and proteins, self-organising maps, disease-gene networks, drug-target networks, and signalling networks [128].

## 5 How to select omic analysis methods

The appropriate tool choice depends on the research question being posed and the study design (Table 7). When creating multiple molecular profiles, an ideal approach is to use the exact same biological sample ("split sample study") [1]. However, the molecular profiles for a sample cannot always be extracted at the exact same timepoints. So, the next best option is biological replicates taken with minimal time differences relative to the source ("replicate-matched") [1]. If testing at the same time is not possible, then at least using the same biological source ("source-matched') can reflect its variation [1]. Finally, if the samples are from equivalent but different biological sources at a varied junctures ("repeated study") then this requires models that include these distinct sources of error [1].

| Omic datasets | | |
|---|---|---|
| **Single** | **Pair** | **Multiple** |
| PCA | Correlation | mCIA |
| MDS | cPCA | NMF extensions |
| CA | CCA | TCA |
| MFA | CIA | WGCNA |
| NMF | PLS-DA* | O2PLS* |
| LDA* | GPA | JIVE |
| | | SNF* |

Table 7. A summary of method classes relevant for analysis of a single omics dataset, or a pair of them, or more than two of them. *Supervised method classes.

Data visualisation, summarisation, quality control and exploration with unsupervised methods are fundamental to most analyses. One useful R package in this respect is dimRed which uses RMSE to compare the distances between the original and transformed values [129]. Another function in the same package, cophenetic_correlation, is the cophenetic correlation between the upper or lower triangles of the distance matrices across the reduced dimensions. DimRed can apply PCA, kPCA, MDS, nonmetric MDS (nMDS), isomap, locally linear embedding (LLE), Laplacian eigenmaps (spectral clustering to separate non-convex clusters), diffusion maps, force directed methods, and dimension reduction via regression (DRR) [130]. The related package coRanking allows comparison of these methods using the relative rank of the data in the original high-dimensional space compared to the new low-dimension one [129]. DRR using kernel ridge regression can be applied with the R package DRR [130]. Diffusion maps can be used via R package diffusionMap [131]. Force directed methods can be implemented with the R package igraph [132].

One common problem with many methods is over-fitting, where a model is trained on sample data that is unrepresentative of the wider population, thus failing to predict well on individuals that were not in the training data [69-70]. This particularly the case for PLS-DA. Cross-validation (CV) can help clarify overfitting in feature selection and classification [133] and is particularly important when applying PLS-DA [64]. CV assesses the generalisability of a model through resampling to estimate the variation in uncertainty across the input data. *K*-fold cross validation is a commonly used method in machine learning to produce estimates of variation within a dataset. The data is split in *K* segments, (typically *K* in 3-10) and a loop of learning is repeated K times. At each iteration one of the segments is left out for the testing, and the remaining *K-1* segments are used for training. Once a new model is trained, the testing segment is used to score how well the trained model performs on this "unseen" data. By producing *K* models and *K* estimates of fitness, we can have a more objective estimate of how well each learning method performs.

Batch error correction should also be considered. Sample isolation, collection and molecular profiling are intensive tasks that take time and typically occur in dispersed labs that possess differing platforms or differentiations of performance of the same platform across sites. Consequently, centre-, technology-, reagent- and technique-associated artefacts are unavailable and can have substantial effects [134]. This can be exacerbated by unbalanced sampling across omic types or study sources that associate group-specific with batch effects so that true effects may be obscured [135]. Numerous approaches can remove unwanted batch effects associated with these known sources of error using empirical Bayes approaches in the R packages sva [136] and ber [137], correction with control genes [138], factor analysis of variation [139], and linear models in R package limma [140]. Another algorithm uses guided PCA to correct across all PCs [141], which has been extended by probabilistic PC and covariates analysis to correct each PC individually for batch effects [142] (the associated packages are no longer available). One way to control for under- or over-correction of batch effects is to repeat the analysis with and without batch effects to quantify the detected effect, or indeed to estimate it with more than one correction tool [135]. Accurate batch effect removal from studies with large design imbalances may be challenging to achieve [143]. ANOVA-simultaneous components analysis (ASCA) differentiates the experiment, batch and residual variation [144]. ASCA removal of systematic noise (ARSyN) is an extension of this: it uses PCA to remove the batch effects for samples shares an omic type [145]. It is available in R package MultiBaC, which applies regression of the shared and unshared information to infer likely missing features in samples with no data for a particular technology, and then applies ARSyN batch effect correction to this data [146].

Consistency and comparability between studies must be improved. This is currently difficult to materialise as there is a lack of databases with the early integration of virus-host multi-omics data. However, there are some resources currently reporting parallel omics results with conceptual integration such as Viruses.String [147] and the multi-omics portal of viral infection (MVIP) [148]. Nonetheless, many databases may be created on an ad hoc basis [149], which makes aligning with findability, accessibility, interoperability and reusability (FAIR) guidance for scientific data management more difficult [150]. Countering this issue, there is a unified query system linking experimental datasets across databases using keywords designed by the European Bioinformatics Institute (EBI) Omics Discovery Index [151]: extending this to individual samples would be useful.

## Conclusions and future prospects

Multi-omic approaches extend the resolution obtained from genomic data to inform on the molecular characteristics of a collection. For example, this could more precisely identify novel changes in virus genomes and host immune responses. This review outlined concepts and challenges in multi-omics, particularly for readers with a limited background in statistics. It outlined the complexity of omic data, how the early integration of such data is pivotal, and a basic background to such integration and dimension reduction. It covered in detail how methods related to PCA, MDS, CA, MFA, NMF and LDA can be used to examine single omic datasets, how cPCA, CCA, CIA, PLS and GPA can be used for pairs of them, and how mCIA, extensions of NMF, TCA, WGCNA, O2PLS, JIVE and SNF can be applied to more than two omic datasets. We discussed multi-omics experiment design, and challenges in data consistency and quality. We detailed how the methods discussed can be directly applied using R, the common language across the majority of researchers.

Numerous challenges remain in this area, stemming from the inherent complexity and heterogeneity of omic data types, coupled with the continuous development of new platforms and sequencing approaches. Some methods have been developed to handle single-cell and spatial transcriptomics data: not only can these be improved, but also how to link these with other omic data types more effectively requires methods that link time-resolved data, spatial information, cellular heterogeneity, cell-cell interactions, and the dynamics of cellular processes. A further related challenge is that surveys show mixed outcomes in terms of alignment of bioinformatics software with FAIR guidelines [152], which may stem from a lack of sufficient training in computing among tool and workflows in biomedical software development [153], including open-source values [154]. At the same time, there are efforts to coordinate multi-omic data processing, quality control and availability [157] and develop sophisticated pan-language methods for multi-omic datasets [158, 159]. It is clear that alignment with FAIR guidance has increasing importance as datasets become more diverse, multi-modal and specific to nuanced experimental design.

## Acknowledgements

The Pirbright Institute receives Institute Strategic Programme Grant funding from the Biotechnology and Biological Sciences Research Council (BBSRC) of the United Kingdom (projects BB/X011038/1, BB/X011046/1, BB/CCG2250/1, BB/IDG2250/1). The funders had no role in the design of the study and collection, analysis, and interpretation of data and in writing the manuscript. The authors thank all anonymous reviewers for their constructive feedback.

## Author Contributions

TD – Conceptualization, Data Curation, Writing - Original Draft, Writing - Review & Editing, Visualisation. NA –Writing - Original Draft, Writing - Review & Editing.

## Data Availability Statement

The data underlying this article are available in the article. The code used to generate the figures is at https://github.com/downingtim/Dimension_reduction_plots/ .

## Competing Interests Statement

The authors declare no competing interests.

**References**

1. Cavill R, et al. Transcriptomic and metabolomic data integration. Brief Bioinform 2016 17(5):891–901.
2. Chauvel C, Novoloaca A, Veyre P, Reynier F, Becker J. Evaluation of integrative clustering methods for the analysis of multi-omics data. Brief Bioinform. 2020 21(2):541-552. doi: 10.1093/bib/bbz015
3. Boulesteix AL, et al. IPF-LASSO: Integrative Penalized Regression with Penalty Factors for Prediction Based on Multi-Omics Data. Comput Math Methods Med. 2017:7691937. doi: 10.1155/2017/7691937.
4. Bodein A, Scott-Boyer MP, Perin O, Lê Cao KA, Droit A. timeOmics: an R package for longitudinal multi-omics data integration. Bioinformatics. 2022 btab664. doi: 10.1093/bioinformatics/btab664.
5. Li CX, Gao J, Zhang Z, Chen L, Li X, Zhou M, Wheelock ÅM. Multiomics integration-based molecular characterizations of COVID-19. Brief Bioinform. 2022 23(1):bbab485. doi: 10.1093/bib/bbab485
6. Reel PS, et al. Using machine learning approaches for multi-omics data analysis: A review. Biotechnology Advances 2021 49:107739. https://doi.org/10.1016/j.biotechadv.2021.107739.
7. Van der Maaten L, Postma E, Van den Herik. Dimensionality Reduction: A Comparative Review. J Mach Learn Res 2009.
8. Paparrizos J, et al. Debunking Four Long-Standing Misconceptions of Time-Series Distance Measures. SIGMOD '20: Proceedings of the 2020 ACM SIGMOD International Conference on Management of DataJune 2020 1887–1905. doi: https://doi.org/10.1145/3318464.3389760
9. Langfelder P, Horvath S. Eigengene networks for studying the relationships between co-expression modules. BMC Syst Biol. 2007 1:54.
10. R Core Team. 2022. R: A language and environment for statistical computing. R Foundation for Statistical Computing, Vienna, Austria. URL https://www.R-project.org
11. Meng C, et al. Dimension reduction techniques for the integrative analysis of multi-omics data. Brief Bioinform. 2016 17(4):628-41. doi: 10.1093/bib/bbv108.
12. Jolliffe IT, Cadima J. Principal component analysis: a review and recent developments. Philos Trans A Math Phys Eng Sci. 2016 374(2065):20150202. doi: 10.1098/rsta.2015.0202.
13. Price AL, et al. Principal components analysis corrects for stratification in genome-wide association studies. Nat. Genet. 2006 38:904–909.
14. Chang CC, Chow CC, Tellier LCAM, Vattikuti S, Purcell SM, Lee JJ. Second-generation PLINK: rising to the challenge of larger and richer datasets. GigaScience 2015 4.
15. Lê Cao KA, Rossouw D, Robert-Granié C, Besse P. A sparse PLS for variable selection when integrating omics data. Stat Appl Genet Mol Biol. 2008 7(1):Article 35. doi: 10.2202/1544-6115.1390.
16. Sigg CD, Buhmann JM. Expectation-Maximization for Sparse and Non-Negative PCA. In Proc. 25th International Conference on Machine Learning 2008 doi: 10.1145/1390156.1390277.
17. Stacklies W, Redestig H, Scholz M, Walther D, Selbig J. PcaMethods -- a Bioconductor package providing PCA methods for incomplete data. Bioinformatics 2007 23:1164-1167.
18. Brouard C, Mariette J, Flamary R, Vialaneix N. Feature selection for kernel methods in systems biology. NAR Genom Bioinform. 2022 4(1):lqac014. doi: 10.1093/nargab/lqac014.
19. Karatzoglou A, Smola A, Hornik K, Zeileis A. kernlab - An S4 Package for Kernel Methods in R. Journal of Statistical Software 2004 11(9):1-20. doi: 10.18637/jss.v011.i09.
20. Le S, Josse J, Husson F. FactoMineR: An R Package for Multivariate Analysis. Journal of Statistical Software 2008 25(1):1-18 doi: 10.18637/jss.v025.i01.
21. Stein-O'Brien GL, et al. Enter the Matrix: Factorization Uncovers Knowledge from Omics. Trends Genet. 2018 34(10):790-805. doi: 10.1016/j.tig.2018.07.003.

22. Lee SI, Batzoglou S. Application of independent component analysis to microarrays. Genome Biol. 2003 4(11):R76. doi: 10.1186/gb-2003-4-11-r76
23. Teschendorff AE, Journée M, Absil PA, Sepulchre R, Caldas C. Elucidating the altered transcriptional programs in breast cancer using independent component analysis. PLoS Comput Biol. 2007 3(8):e161. doi: 10.1371/journal.pcbi.0030161.
24. Engreitz JM, Daigle BJ Jr, Marshall JJ, Altman RB. Independent component analysis: mining microarray data for fundamental human gene expression modules. J Biomed Inform. 2010 43(6):932-44. doi: 10.1016/j.jbi.2010.07.001.
25. Lê Cao KA, Boitard S, Besse P. Sparse PLS discriminant analysis: biologically relevant feature selection and graphical displays for multiclass problems. BMC Bioinformatics. 2011 12:253. doi: 10.1186/1471-2105-12-253.
26. Bartenhagen C, Klein HU, Ruckert C, Jiang X, Dugas M. Comparative study of unsupervised dimension reduction techniques for the visualization of microarray gene expression data. BMC Bioinformatics. 2010 11:567. doi: 10.1186/1471-2105-11-567
27. Tenenbaum JB, de Silva V, Langford JC. A global geometric framework for nonlinear dimensionality reduction. Science. 2000 290(5500):2319-23. doi: 10.1126/science.290.5500.2319.
28. Lee DD, Seung HS. Learning the parts of objects by non-negative matrix factorization. Nature. 1999 401(6755):788-91. doi: 10.1038/44565.
29. Kim PM, Tidor B. Subsystem identification through dimensionality reduction of large-scale gene expression data. Genome Res. 2003 13(7):1706-18. doi: 10.1101/gr.903503.
30. Devarajan K. Nonnegative matrix factorization: an analytical and interpretive tool in computational biology. PLoS Comput Biol. 2008 4(7):e1000029. doi: 10.1371/journal.pcbi.1000029.
31. Brunet JP, Tamayo P, Golub TR, Mesirov JP. Metagenes and molecular pattern discovery using matrix factorization. Proc Natl Acad Sci U S A. 2004 101(12):4164-9. doi: 10.1073/pnas.0308531101.
32. Ochs MF, Fertig EJ. Matrix factorization for transcriptional regulatory network inference. IEEE Symp. Comput. Intell. Bioinforma. Comput. Biol. Proc. 2012 17:387–396.
33. Kossenkov AV, et al. Determining transcription factor activity from microarray data using Bayesian Markov chain Monte Carlo sampling. Stud. Health Technol. Inform. 2007 129:1250–1254.
34. Mairal J, et al. Online learning for matrix factorization and sparse coding. J. Mach. Learn. Res. 2010 11:19–60.
35. Gaujoux R, Seoighe C. A flexible R package for nonnegative matrix factorization. BMC Bioinformatics. 2010 11:367. doi: 10.1186/1471-2105-11-367
36. Fertig EJ, Ding J, Favorov AV, Parmigiani G, Ochs MF. CoGAPS: an integrated R/C++ package to identify overlapping patterns of activation of biological processes from expression data. Bioinformatics 2010 26(21), 2792–2793.
37. Dray S, Chessel D, Thioulouse J . Co-inertia analysis and the linking of ecological data tables. Ecology 2003 84:3078–89.
38. Venables WN, Ripley BD. Modern Applied Statistics with S. Fourth Edition. Springer, New York. 2002. ISBN 0-387-95457-0.
39. Nenadic O, Greenacre, M. Correspondence Analysis in R, with two- and three-dimensional graphics: The ca package. Journal of Statistical Software 2007 20(3):1-13.
40. Beh EJ, Lombardo R. A genealogy of correspondence analysis. Australian & New Zealand Journal of Statistics 2012 54(2):137-168.
41. Dray S, Dufour A. The ade4 Package: Implementing the Duality Diagram for Ecologists. Journal of Statistical Software 2007 22(4):1-20. doi:10.18637/jss.v022.i04.
42. De Tayrac M, Lê S, Aubry M, Mosser J, Husson F. Simultaneous analysis of distinct Omics data sets with integration of biological knowledge: Multiple Factor Analysis approach. BMC Genomics 2009 10:32. doi: 10.1186/1471-2164-10-32.

43. Westerhuis JA. Kourti T MacGregor JF . Analysis of multiblock and hierarchical PCA and PLS models . J Chemom 1998 12:301–21.
44. Hassani S, Martens H, Qannari E, Hanafi M, Borge G, Kohler A. Analysis of omics data: graphical interpretation- and validation tools in multi-block methods. Chemomet Intel Lab Syst. 2010 104:140–153.
45. Conesa A, Prats-Montalbán J, Tarazona S, Nueda MJ, Ferrer A. A multiway approach to data integration in systems biology based on Tucker3 and N-PLS. Chemomet Intel Lab Syst. 2010 104:101–111.
46. Meng C. mogsa: Multiple omics data integrative clustering and gene set analysis. 2021. R package version 1.26.0.
47. Hotelling H. Relations between two sets of variates. Biometrika. 1936; 28(3/4):321 https://doi.org/10.2307/2333955.
48. Hong S, Chen X, Jin L, Xiong M. Canonical correlation analysis for RNA-seq co-expression networks. Nucleic Acids Res. 2013 41(8):e95. doi: 10.1093/nar/gkt145.
49. Witten D, Tibshirani R. Extensions of sparse canonical correlation analysis with applications to genomic data. Statistical Applications in Genetics and Molecular Biology. 2009 8(1):28.
50. Witten DM, Tibshirani R, Hastie T. A penalized matrix decomposition, with applications to sparse principal components and canonical correlation analysis. Biostatistics. 2009 10(3):515-34. doi: 10.1093/biostatistics/kxp008
51. Nguyen T, Tagett R, Diaz D, Draghici S. A novel approach for data integration and disease subtyping. Genome Res. 2017 27(12):2025-2039. doi: 10.1101/gr.215129.116
52. Shi WJ, et al. Unsupervised discovery of phenotype-specific multi-omics networks. Bioinformatics. 2019 35(21):4336-4343. doi: 10.1093/bioinformatics/btz226.
53. Chessel D, Doledec S. Co-inertia analysis: an alternative method for studying species–environment relationships. Freshwater Biology 1994 31(3): 277-294 doi: https://doi.org/10.1111/j.1365-2427.1994.tb01741.x.
54. Chessel D, Hanafi M. Analyses de la co-inertie de k nuages de points. Revue de statistique appliquée. 1996 44(2):35–60.
55. Wu C, Zhou F, Ren J, Li X, Jiang Y, Ma S. A Selective Review of Multi-Level Omics Data Integration Using Variable Selection. High Throughput. 2019 8(1):4. doi: 10.3390/ht8010004
56. Min EJ, Safo SE, Long Q. Penalized co-inertia analysis with applications to -omics data. Bioinformatics. 2019 35(6):1018-1025. doi: 10.1093/bioinformatics/bty726.
57. Culhane AC, Perrière G, Higgins DG. Cross-platform comparison and visualisation of gene expression data using co-inertia analysis. BMC Bioinformatics. 2003 4:59. doi: 10.1186/1471-2105-4-59.
58. Stanimirova I, et al. STATIS, a three-way method for data analysis. Application to environmental data. Chemometrics & Intelligent Laboratory Systems 2004 73(2):219-233 doi: https://doi.org/10.1016/j.chemolab.2004.03.005.
59. Mendes S, et al. The efficiency of the Partial Triadic Analysis method: an ecological application. Biometrical Letters 2010 47(2):83-106.
60. Zang C, et al. High-dimensional genomic data bias correction and data integration using MANCIE. Nat Commun. 2016 7:11305. doi: 10.1038/ncomms11305
61. Lock EF, Hoadley KA, Marron JS, Nobel AB. Joint and individual variation explained (JIVE) for integrated analysis of multiple data types. Ann Appl Stat. 2013 7(1):523-542. doi: 10.1214/12-AOAS597.
62. Liquet B, De Micheaux PL, Hejblum BP, Thiébaut R. Group and sparse group partial least square approaches applied in genomics context. Bioinformatics. 2016 32(1):35–42. https://doi.org/10.1093/bioinformatics/btv535.
63. Barker M, Rayens W. Partial least squares for discrimination. J Chemom. 2003 17(3):166–73.

64. Ruiz-Perez D, Guan H, Madhivanan P, Mathee K, Narasimhan G. So you think you can PLS-DA? BMC Bioinformatics. 2020 21(Suppl 1):2. doi: 10.1186/s12859-019-3310-7.
65. Shaw CD, et al. Genomic and Metabolomic Polymorphism among Experimentally Selected Paromomycin-Resistant Leishmania donovani Strains. Antimicrob Agents Chemother. 2019 64(1):e00904-19. doi: 10.1128/AAC.00904-19
66. Kuhn M. caret: Classification and Regression Training. R package version 6.0-92. 2022. https://CRAN.R-project.org/package=caret.
67. Eslami A, Qannari EM, Kohler A, Bougeard S. Algorithms for multi-group PLS. J Chemometrics. 2014 28(3):192–201.
68. Rohart F, Eslami A, Matigian N, Bougeard S, Lê Cao KA. MINT: a multivariate integrative method to identify reproducible molecular signatures across independent experiments and platforms. BMC Bioinformatics. 2017 18(1):128. doi: 10.1186/s12859-017-1553-8.
69. Brereton RG, Lloyd GR. Partial least squares discriminant analysis: taking the magic away. Journal Chemometrics 28(4):213-225 2014 doi: https://doi.org/10.1002/cem.2609
70. Rodríguez-Pérez R, Fernández L, Marco S. Overoptimism in cross-validation when using partial least squares-discriminant analysis for omics data: a systematic study. Anal Bioanal Chem. 2018 410(23):5981-5992. doi: 10.1007/s00216-018-1217-1
71. Goodall C. Procrustes methods in the statistical analysis of shape. Journal of the Royal Statistical Society B 1991 53(2):285–339.
72. Pagès J. Multiple Factor Analysis by Example Using R. 2014. Chapman & Hall/CRC The R Series, London.
73. Abdi H, Williams LJ, Valentin D. Multiple factor analysis: principal component analysis for multitable and multiblock data sets. WIREs Comp Stat. 2013 5:149–79. doi: 10.1002/wics.1246
74. Oksanen J, et al. vegan: Community Ecology Package. R package version 2.6-2. 2022. https://CRAN.R-project.org/package=vegan.
75. Meng C, Kuster B, Culhane AC, Gholami AM. A multivariate approach to the integration of multi-omics datasets. BMC Bioinformatics. 2014 15:162. doi: 10.1186/1471-2105-15-162.
76. Min EJ, Long Q. Sparse multiple co-Inertia analysis with application to integrative analysis of multi -Omics data. BMC Bioinformatics. 2020 21(1):141. doi: 10.1186/s12859-020-3455-4.
77. O'Connell MJ, Lock EF. R.JIVE for exploration of multi-source molecular data. Bioinformatics. 2016 32(18):2877-9. doi: 10.1093/bioinformatics/btw324.
78. Hellton KH, Thoresen M. Integrative clustering of high-dimensional data with joint and individual clusters. Biostatistics. 2016 17(3):537-48. doi: 10.1093/biostatistics/kxw005.
79. Yang X, Hoadley KA, Hannig J, Marron JS. 2023. Jackstraw inference for AJIVE data integration. Computational Statistics & Data Analysis 180:107649 doi: 10.1016/j.csda.2022.107649.
80. Carmichael I. 2020. idc9/r_jive: First github release doi: 10.5281/zenodo.4091755
81. Brown BC, et al. Multiset correlation and factor analysis enables exploration of multi-omic data. BioRxiv 2022 doi: https://doi.org/10.1101/2022.07.18.500246.
82. Tenenhaus A, Tenenhaus M. Regularized generalized canonical correlation analysis for multiblock or multigroup data analysis. Eur J Oper Res. 2014; 238(2):391–403.
83. Tenenhaus A, Philippe C, Guillemot V, Le Cao KA, Grill J, Frouin V. Variable selection for generalized canonical correlation analysis. Biostatistics. 2014 15(3):569-83. doi: 10.1093/biostatistics/kxu001
84. Luo Y, Tao D, Ramamohanarao K, Xu C, Wen Y. Tensor canonical correlation analysis for multi-view dimension reduction. Proc. ICDE 2016. 2016 1460–1461.
85. Min W, Chang TH, Zhang S, Wan X. TSCCA: A tensor sparse CCA method for detecting microRNA-gene patterns from multiple cancers. PLoS Comput Biol. 2021 17(6):e1009044. doi: 10.1371/journal.pcbi.1009044.

86. Singh A, Shannon CP, Gautier B, Rohart F, Vacher M, Tebbutt SJ, Lê Cao KA. DIABLO: an integrative approach for identifying key molecular drivers from multi-omics assays. Bioinformatics. 2019 35(17):3055-3062. doi: 10.1093/bioinformatics/bty1054
87. Carroll JD, Chang JJ. Analysis of individual differences in multidimensional scaling via an n-way generalization of "Eckart–young" decomposition. Psychometrika 1970 35:283–319. doi: 10.1007/BF02310791.
88. Harshman RA, Lundy ME. PARAFAC: Parallel factor analysis. Computational Statistics & Data Analysis 1999 18(1):39-72.
89. Kiers HAL. A three–step algorithm for CANDECOMP/PARAFAC analysis of large data sets with multicollinearity. Journal of Chemometrics 1999 12(3):155-171 doi https://doi.org/10.1002/(SICI)1099-128X(199805/06)12:3<155::AID-CEM502>3.0.CO;2-5.
90. Qiao X, Zhang X, Chen W, Xu X, Chen YW, Liu ZP. tensorGSEA: Detecting Differential Pathways in Type 2 Diabetes via Tensor-Based Data Reconstruction. Interdiscip Sci. 2022 14(2):520-531. doi: 10.1007/s12539-022-00506-2.
91. Taguchi YH. Tensor decomposition-based unsupervised feature extraction applied to matrix products for multi-view data processing. PLoS One. 2017 12(8):e0183933. doi: 10.1371/journal.pone.0183933
92. Tucker LR. Some mathematical notes on three-mode factor analysis. Psychometrika. 1966 31(3):279-311. doi: 10.1007/BF02289464.
93. Yang J, Zhang D, Frangi AF, Yang JY. Two-dimensional PCA: a new approach to appearance-based face representation and recognition. IEEE Trans Pattern Anal Mach Intell. 2004 26(1):131-7. doi: 10.1109/tpami.2004.1261097
94. Kong H, et al. Generalized 2D principal component analysis for face image representation and recognition. Neural Netw 2005 18:585–594. https://doi.org/10.1016/j.neunet.2005.06.041.
95. Yu H, Bennamoun M. 1D-PCA, 2D-PCA to nD-PCA. 2006. 18th International Conference on Pattern Recognition (ICPR'06) 181-184 doi: 10.1109/ICPR.2006.19.
96. Hore V, et al. Tensor decomposition for multiple-tissue gene expression experiments. Nat Genet 2016 48:1094–1100. https://doi.org/10.1038/ng.3624.
97. Marchini JL, Heaton C, Ripley BD. fastICA: FastICA Algorithms to Perform ICA and Projection Pursuit. R package version 1.2-3. 2021 https://CRAN.R-project.org/package=fastICA.
98. Mor U, Cohen Y, Valdés-Mas R, Kviatcovsky D, Elinav E, Avron H. Dimensionality reduction of longitudinal 'omics data using modern tensor factorizations. PLoS Comput Biol. 2022 18(7):e1010212. doi: 10.1371/journal.pcbi.1010212.
99. Jung I, Kim M, Rhee S, Lim S, Kim S. MONTI: A Multi-Omics Non-negative Tensor Decomposition Framework for Gene-Level Integrative Analysis. Front Genet. 2021 12:682841. doi: 10.3389/fgene.2021.682841.
100. Leibovici DG. Spatio-Temporal Multiway Decompositions Using Principal Tensor Analysis on k-Modes: The R Package PTAk. Journal of Statistical Software 2010 34(10):1-34. doi:10.18637/jss.v034.i10.
101. Giordani P, Kiers HAL, Del Ferraro MA. Three-Way Component Analysis Using the R Package ThreeWay. Journal of Statistical Software 2014 57(7):1-23. doi: http://www.jstatsoft.org/v57/i07/.
102. Virta J, Koesner CL, Li B, Nordhausen K, Oja H, Radojicic U. tensorBSS: Blind Source Separation Methods for Tensor-Valued Observations. 2021. R package version 0.3.8. https://CRAN.R-project.org/package=tensorBSS.
103. Li J, Bien J, Wells MT. rTensor: An R Package for Multidimensional Array (Tensor) Unfolding, Multiplication, and Decomposition. Journal of Statistical Software 2018 87(10):1–31. https://doi.org/10.18637/jss.v087.i10.
104. Allaire JJ, Tang Y. tensorflow: R Interface to 'TensorFlow'. R package version 2.9.0. 2022. https://CRAN.R-project.org/package=tensorflow.

105. Zamora R. tensorr: Sparse Tensors in R. R package version 0.1.1. 2019. https://CRAN.R-project.org/package=tensor.
106. Gill CC, Marchini J. Four-Dimensional Sparse Bayesian Tensor Decomposition for Gene Expression Data 2020 BioRxiv doi: https://doi.org/10.1101/2020.11.30.403907.
107. Trygg J, Wold S. O2-PLS, a two-block (X-Y) latent variable regression (LVR) method with an integral OSC filter. Journal of Chemometrics. 2003 17:53–64.
108. Gill CC, Marchini J. Four-Dimensional Sparse Bayesian Tensor Decomposition for Gene Expression Data 2020 BioRxiv doi: https://doi.org/10.1101/2020.11.30.403907.
109. Gu Z, El Bouhaddani S, Pei J, Houwing-Duistermaat J, Uh HW. Statistical integration of two omics datasets using GO2PLS. BMC Bioinformatics. 2021 22(1):131. doi: 10.1186/s12859-021-03958-3.
110. Löfstedt T, Hoffman D, Trygg J. Global, local and unique decompositions in OnPLS for multiblock data analysis. Anal Chim Acta. 2013 791:13-24. doi: 10.1016/j.aca.2013.06.026.
111. Reinke SN, et al. OnPLS-Based Multi-Block Data Integration: A Multivariate Approach to Interrogating Biological Interactions in Asthma. Anal Chem. 2018 90(22):13400-13408. doi: 10.1021/acs.analchem.8b03205.
112. Li W, Zhang S, Liu CC, Zhou XJ. Identifying multi-layer gene regulatory modules from multi-dimensional genomic data. Bioinformatics. 2012 28(19):2458-66. doi: 10.1093/bioinformatics/bts476
113. Bersanelli M, et al. Methods for the integration of multi-omics data: mathematical aspects. BMC Bioinformatics. 2016 17 Suppl 2(Suppl 2):15. doi: 10.1186/s12859-015-0857-9.
114. Tini G, Marchetti L, Priami C, Scott-Boyer MP. Multi-omics integration-a comparison of unsupervised clustering methodologies. Brief Bioinform. 2019 20(4):1269-1279. doi: 10.1093/bib/bbx167.
115. Wang B, et al. Similarity network fusion for aggregating data types on a genomic scale. Nat Methods. 2014 11(3):333-7. doi: 10.1038/nmeth.2810
116. Wang B, et al. SNFtool: Similarity Network Fusion. R package version 2.3.1. 2021. https://CRAN.R-project.org/package=SNFtool
117. Langfelder P, Horvath S. WGCNA: an R package for weighted correlation network analysis. BMC Bioinformatics. 2008 9:559. doi: 10.1186/1471-2105-9-559.
118. Valdeolivas A, et al. Random walk with restart on multiplex and heterogeneous biological networks. Bioinformatics. 2019 35(3):497-505. doi: 10.1093/bioinformatics/bty637
119. Decano AG, et al. Plasmids shape the diverse accessory resistomes of Escherichia coli ST131. Access Microbiol. 2020 3(1):acmi000179. doi: 10.1099/acmi.0.000179.
120. Nicolau M, Levine AJ, Carlsson G. Topology based data analysis identifies a subgroup of breast cancers with a unique mutational profile and excellent survival. Proc Natl Acad Sci U S A. 2011 108(17):7265-70. doi: 10.1073/pnas.1102826108.
121. Downing T, Rahm A. Bacterial plasmid-associated and chromosomal proteins have fundamentally different properties in protein interaction networks. Sci Rep. 2022 12(1):19203. doi: 10.1038/s41598-022-20809-0
122. Downing T, Lee MJ, Archbold C, McDonnell A, Rahm A. Informing plasmid compatibility with bacterial hosts using protein-protein interaction data. Genomics. 2022 114(6):110509. doi: 10.1016/j.ygeno.2022.110509
123. Vietoris L. Über den höheren Zusammenhang kompakter Räume und eine Klasse von zusammenhangstreuen Abbildungen. Mathematische Annalen 1927 97:454–472.
124. Cámara PG. Topological methods for genomics: present and future directions. Curr Opin Syst Biol. 2017 1:95-101. doi: 10.1016/j.coisb.2016.12.007.
125. Argelaguet R, et al. Multi-Omics Factor Analysis-a framework for unsupervised integration of multi-omics data sets. Mol Syst Biol. 2018 14(6):e8124. doi: 10.15252/msb.20178124
126. Kirk P, Griffin JE, Savage RS, Ghahramani Z, Wild DL. Bayesian correlated clustering to integrate multiple datasets. Bioinformatics. 2012 28(24):3290-7. doi: 10.1093/bioinformatics/bts595

127. Lock EF, Dunson DB. Bayesian consensus clustering. Bioinformatics. 2013 29(20):2610-6. doi: 10.1093/bioinformatics/btt425
128. Kamburov A, Cavill R, Ebbels TM, Herwig R, Keun HC. Integrated pathway-level analysis of transcriptomics and metabolomics data with IMPaLA. Bioinformatics. 2011 27(20):2917-8. doi: 10.1093/bioinformatics/btr499.
129. Kraemer G, Reichstein M, Mahecha MD. dimRed and coRanking-Unifying Dimensionality Reduction in R. The R Journal 2018 10(1):342-358 doi: https://journal.r-project.org/archive/2018/RJ-2018-039/
130. Kraemer G. DRR: Dimensionality Reduction via Regression. R package version 0.0.4. 2020 https://CRAN.R-project.org/package=DRR.
131. Richards J, Cannoodt R. diffusionMap: Diffusion Map. R package version 1.2.0. 2019. https://CRAN.R-project.org/package=diffusionMap.
132. Csardi G, Nepusz T. The igraph software package for complex network research. InterJournal, Complex Systems 2006 1695. https://igraph.org.
133. Westerhuis JA, Hoefsloot HC, Smit S, Vis DJ, Smilde AK, van Velzen EJ, van Duijnhoven JP, van Dorsten FA. Assessment of PLSDA cross validation. Metabolomics. 2008 4(1):81–9.
134. Kupfer P, et al. Batch correction of microarray data substantially improves the identification of genes differentially expressed in rheumatoid arthritis and osteoarthritis. BMC Med. Genomics 2012 5:1–12.
135. Nygaard V, Rødland EA, Hovig E. Methods that remove batch effects while retaining group differences may lead to exaggerated confidence in downstream analyses. Biostatistics. 2016 17(1):29-39. doi: 10.1093/biostatistics/kxv027
136. Leek JT, et al. The SVA package for removing batch effects and other unwanted variation in high-throughput experiments. Bioinformatics 2012 28:882–883.
137. Giordan M. A two-stage procedure for the removal of batch effects in microarray studies. Stat Biosci 2014 6(1):73–84.
138. Gagnon-Bartsch JA, Speed TP. Using control genes to correct for unwanted variation in microarray data. Biostatistics 2012 13(3): 539–552.
139. Risso D, et al. Normalization of RNA-seq data using factor analysis of control genes or samples. Nat. Biotechnol. 2014 32:896–902.
140. Ritchie ME, et al. limma powers differential expression analyses for RNA-sequencing and microarray studies. Nucleic Acids Research 2015 43:e47.
141. Reese SE, et al. A new statistic for identifying batch effects in high-throughput genomic data that uses guided principal component analysis. Bioinformatics 2013 29(22): 2877–2883.
142. Nyamundanda G, Poudel P, Patil Y, Sadanandam A. A Novel Statistical Method to Diagnose, Quantify and Correct Batch Effects in Genomic Studies. Sci Rep. 2017 (1):10849. doi: 10.1038/s41598-017-11110-6
143. Tarazona S, et al. Harmonization of quality metrics and power calculation in multi-omic studies. Nat. Commun. 2020 11:3092.
144. Jansen JJ, et al. ASCA: Analysis of multivariate data obtained from an experimental design. J Chemom 2005 19(9): 469–481.
145. Nueda MJ, et al. ARSyN: a method for the identification and removal of systematic noise in multifactorial time course microarray experiments. Biostatistics 2012 13:553–566.
146. Ugidos M, et al. MultiBaC: A strategy to remove batch effects between different omic data types. Stat Methods Med Res. 2020 29(10):2851-2864. doi: 10.1177/0962280220907365
147. Cook HV, Doncheva NT, Szklarczyk D, von Mering C, Jensen LJ. Viruses.STRING: A Virus-Host Protein-Protein Interaction Database. Viruses. 2018 10(10):519. doi: 10.3390/v10100519
148. Tang Z, Fan W, Li Q, Wang D, Wen M, Wang J, Li X, Zhou Y. MVIP: multi-omics portal of viral infection. Nucleic Acids Res. 2022 50(D1):D817-D827. doi: 10.1093/nar/gkab958
149. Tarazona, S, et al. Undisclosed, unmet and neglected challenges in multi-omics studies. Nat Comput Sci 2021 1:395–402 doi: https://doi.org/10.1038/s43588-021-00086-z

150. Wilkinson M, et al. The FAIR guiding principles for scientific data management and stewardship, Scientific Data 2016 3:160018. doi:10.1038/sdata.2016.18.
151. Perez-Riverol Y, et al. Discovering and linking public omics data sets using the Omics Discovery Index. Nat Biotechnol. 2017 35(5):406-409. doi: 10.1038/nbt.3790
152. Russell PH, et al. A large-scale analysis of bioinformatics code on GitHub. PLoS One. 2018 13(10):e0205898. doi: 10.1371/journal.pone.0205898.
153. Silva LB, Jimenez RC, Blomberg N, Luis Oliveira J. General guidelines for biomedical software development. F1000Res. 2017 6:273. doi: 10.12688/f1000research.10750.2
154. Jiménez RC, et al. Four simple recommendations to encourage best practices in research software. F1000Res. 2017 6:ELIXIR-876. doi: 10.12688/f1000research.11407.1
155. González I, Déjean S. CCA: Canonical Correlation Analysis. 2021. R package version 1.2.1. https://CRAN.R-project.org/package=CCA.
156. Bodein A, Scott-Boyer MP, Perin O, Lê Cao KA, Droit A. Interpretation of network-based integration from multi-omics longitudinal data. Nucleic Acids Res. 2022 50(5):e27. doi: 10.1093/nar/gkab1200.
157. Conesa A, Beck S. Making multi-omics data accessible to researchers. Sci Data. 2019 6(1):251. doi: https://doi.org/10.1038/s41597-019-0258-4.
158. Bredikhin D, Kats I, Stegle O. MUON: multimodal omics analysis framework. Genome Biol. 2022 23(1):42. doi: 10.1186/s13059-021-02577-8
159. Vandereyken K, Sifrim A, Thienpont B, Voet T. Methods and applications for single-cell and spatial multi-omics. Nat Rev Genet. 2023 24(8):494-515. doi: 10.1038/s41576-023-00580-2